\documentclass[journal]{IEEEtran}
\ifCLASSINFOpdf
\else
\fi
\usepackage{amsmath}
\usepackage{amsfonts}
\usepackage{enumerate}
\usepackage{amssymb}
\usepackage{psfrag}
\usepackage{color}
\usepackage{cite}
\newtheorem{theorem}{Theorem}

\newtheorem{corollary}{Corollary}

\newtheorem{definition}{Definition}
\newtheorem{property}{Property}

\makeatletter
\newcommand{\smatlabaxislabel}[1]{\fontsize{12}{\f@baselineskip}%
\textsf{#1}}
\newcommand{\matlabaxislabel}[1]{\fontsize{14.4}{\f@baselineskip}%
\textsf{#1}}
\newcommand{\mmatlabaxislabel}[1]{\fontsize{17.28}{\f@baselineskip}%
\textsf{#1}}
\newcommand{\bmatlabaxislabel}[1]{\fontsize{20.74}{\f@baselineskip}%
\textsf{#1}}
\newcommand{\bbmatlabaxislabel}[1]{\fontsize{24.88}{\f@baselineskip}%
\textsf{#1}} \makeatother

\allowdisplaybreaks

\begin{document}

\title{Full
Diversity Codes for MISO Systems Equipped with Linear or ML
Detectors}

\author{Jing~Liu,~\IEEEmembership{Student Member, IEEE,}
        Jian-Kang~Zhang,~\IEEEmembership{Member, IEEE,}
        and~Kon~Max~Wong,~\IEEEmembership{Fellow,~IEEE}
\thanks{The authors are with the Department of
Electrical and Computer Engineering, McMaster University, Hamilton,
Ontario, Canada e-mails: jliu@grads.ece.mcmaster.ca, (jkzhang,
wong)@mail.ece.mcmaster.ca}}

\markboth{Journal of IEEE Trans. on Information Theory,~Vol.~XX, No.~X, XXX~200X}%
{Shell \MakeLowercase{\textit{et al.}}: Bare Demo of IEEEtran.cls for Journals}

\maketitle

\begin{abstract}
In this paper, a general criterion for space time block codes (STBC)
to achieve full-diversity with a linear receiver is proposed for a
wireless communication system having multiple transmitter and single
receiver antennas (MISO). Particularly, the STBC with Toeplitz
structure satisfies this criterion and therefore, enables
full-diversity. Further examination of this Toeplitz STBC reveals
the following important properties: a) The symbol transmission rate
can be made to approach unity. b) Applying the Toeplitz code to any
signalling scheme having nonzero distance between the nearest
constellation points results in a non-vanishing determinant. In
addition, if QAM is used as the signalling scheme, then for
independent MISO flat fading channels, the Toeplitz codes is proved
to approach the optimal diversity-vs-multiplexing tradeoff with a ZF
receiver when the number of channel uses is large. This is, so far,
the first non-orthogonal STBC shown to achieve the optimal tradeoff
for such a receiver. On the other hand, when ML detection is
employed in a MISO system, the Toeplitz STBC achieves the maximum
coding gain for independent channels. When the channel fading
coefficients are correlated, the inherent transmission matrix in the
Toeplitz STBC can be designed to minimize the average worst case
pair-wise error probability.
\end{abstract}


\section{Introduction}\label{sec:Introduction}
\IEEEPARstart{T}{he} recent arrival of the Information Age has
created an explosive demand for knowledge and information exchange
in our society. This demand has triggered off an enormous expansion
in wireless communications in which severe technical challenges,
including the need of transmitting speech, data and video at high
rates in an environment rich of scattering, have been encountered. A
recent development in wireless communication systems is the
multi-input multi-output (MIMO) wireless link which, due to its
potential in meeting these challenges caused by fading channels
together with power and bandwidth limitations, has become a very
important area of research. The importance of MIMO communications
lies in the fact that they are able to provide a significant
increase in capacity over single-input single-output (SISO)
channels. Existing MIMO designs employ multiple transmitter antennas
and multiple receiver antennas to exploit the high symbol rate
provided by the capacity available in the MIMO channels. Full symbol
rate is achieved when, on average, one symbol is transmitted by each
of the multiple transmitter antennas per time slot (often called a
``channel use"). In the case of $M$ transmitter antennas, we will
have an average of $M$ symbols per channel use (pcu) at full rate.
Furthermore, to combat fading and cross-talk, MIMO systems provide
different replicas of transmitted symbols to the receiver by using
multiple receiver antennas with sufficient separation between each
so that the fading for the receivers are independent of each other.
Such diversity can also be achieved at the transmitter by spacing
the transmitter antennas sufficiently and introducing a code for the
transmitted symbols distributed over transmitter antennas (space)
and symbol periods (time), i.e., space-time
coding~\cite{guey96,alamouti98,tarokh99, damen02}. Full diversity is
achieved when the total degree of freedom available in the
multi-antenna system is utilized.

Over the past several years, various space-time coding schemes have
been developed to take advantage of the MIMO communication channel.
Using a linear processor, orthogonal space-time block
codes~\cite{geramita79,alamouti98,tarokh99,ganesan01,tirkkonen02,liang03}
can provide maximum diversity achievable by a maximum likelihood
detector. However, they have a limited transmission
rate~\cite{liang01,su01,wang02,liang03} and thus, do not achieve
full MIMO channel capacity~\cite{sandhu00}. Linear dispersion codes
have been proposed in~\cite{hassibi02} for which each transmitted
codeword is a linear combination of certain weighted matrices
maximizing the ergodic capacity of the system. Unfortunately, good
error probability performance for these codes is not strictly
guaranteed. To bridge the gap between multiplexing and diversity, a
linear dispersion code design has been proposed using frame
theory~\cite{heath02} that typically performs well both in terms of
ergodic capacity and error performance, but full diversity still
cannot be guaranteed. Thus far, with the exception of the orthogonal
STBC, all existing STBC are designed such that full diversity can
only be achieved when the ML detector is employed. Recent
research~\cite{gamal03, ma03, sethuraman03, jkz-icassp04,
jkz-isit04} based on number theory has shown that employing a ML
receiver, it is possible to design linear space-time block codes and
dispersion codes which are full rate and full diversity without
information loss. The major concern on these designs is that the
coding gain vanishes rapidly as the constellation size increases.
Therefore, designs of full-rate, full-diversity space-time codes
with non-vanishing coding gain have drawn much
attention~\cite{yao03, dayal03, belfiore03, belfiore04, rekaya04,
wang-isit04,jkz-icassp05,gyw-icassp05,gamal04,dense, rajan05,
kumar2006, golden2006}
since 
such structured space-time codes could achieve the optimal
diversity-vs-multiplexing tradeoff developed by Zheng and
Tse\cite{zheng03}. However, most available STBC possessing these
properties are for ML receivers only.

In this paper, we consider a coherent communication system equipped
with multiple transmitter antennas and a single receiver antenna,
i.e., a MISO system. These systems are often employed in mobile
communications for which the mobile receiver may not be able to
support multiple antennas. The highest transmission rate for a MISO
system is unity, i.e., one symbol pcu. For such a MISO system with
ML receivers, rate-1 and full diversity STBC have been proposed by
various authors~\cite{quasi-xia, quasi-sharma, quasi-dao,dense}. In
this paper, however, we consider such a MISO system equipped with
\emph{linear receivers} for which we propose a general criterion for
the design of a full-diversity STBC. In particular, we introduce the
Toeplitz STBC as a member of the family of the full diversity STBC.
It should be noted that the Toeplitz structure has already been
successfully employed as a special case of the delay diversity code
(DDC) \cite{Paulraj-DCC,winter-DCC,Gore-book93,winter94} applied to
MIMO systems having outer channel coding and ML detection. Here, we
extend its application to the construction of STBC in a MISO system
by having a Toeplitz coding matrix cascaded 
with a beamforming matrix. We show that the Toeplitz STBC has
several important properties which enable the code, when applied to
a MISO system with a linear receiver, to asymptotically achieve unit
symbol rate, to possess non-vanishing determinants for signal
constellations having non-zero distance between nearest neighbours,
and to achieve full diversity~\cite{zlw-isit05} accomplishing the
optimal tradeoff of diversity and multiplexing gains~\cite{zheng03}.

On the other hand, we also consider the MISO system in which the
channel has zero mean and fixed covariance known to the transmitter.
For such MISO systems, sacrificing the transmission rate by
repeating the transmitted symbols, and employing maximum ratio
combining together with orthogonal space-time coding, an optimal
precoder can be designed~\cite{zhou-sp02,zhou-it03} by minimizing
the upper bound of the average symbol error probability (SEP). Here
in this paper, we apply the Toeplitz STBC to such a MISO system.
Maintaining rate one and full diversity, we present a design that
minimizes the \emph{exact} worst case average pair-wise error
probability when the ML detector is employed at the receiver.

\section{MISO System Model and Properties of the
Channel Matrix}\label{sec:DesignCriterion}

Consider a MIMO communication system having $M$ transmitter antennas
and $M_R$ receiver antennas transmitting the symbols
$\{s_\ell\},~ \ell = 1, \ldots, L$ which are selected from a given
constellation, i.e., $s_\ell\in \mathcal S$.
To facilitate the transmission of these $L$ symbols through the $M$
antennas in the $N$ time slots (channel use), each symbol is
processed by an $N\times M$ coding matrix $\mathbf{A}_\ell$, 
and then summed together, resulting in an $N\times M$ STBC matrix
given by
    $\mathbf{X}=\sum_{\ell=1}^{L}s_\ell\mathbf{A}_\ell$
where the ($nm$)th element of $\mathbf{X}$ represents the coded
symbol to be transmitted from the $m$th antenna at the $n$th time
slot. These coded symbols are then transmitted to the receiver
antennas through flat-fading path coefficients which form the
elements of the $M \times M_R$ channel matrix $\mathbf{H}$. 
The received space-time signal, denoted by the $N \times M_R$ matrix
$\mathbf{Y}$, can be written as
\begin{equation}\label{eq:MIMOsystem}
    \mathbf{Y}=     
    \mathbf{XH}+\boldsymbol{\Xi}
\end{equation}
where 
$\boldsymbol{\Xi}$ is the $N \times M_R$ additive white space-time
noise matrix whose elements are of complex circular Gaussian
distribution $\mathcal{CN}(0,1)$.


Let us now turn our attention to a MISO wireless communication
system which is a special case of the MIMO system having $M$
transmitter antennas and a single receiver antenna. Just as in the
MIMO system, the transmitted symbols $s_\ell$, $\ell=1,\cdots,L$ in
the MISO system are coded by linear $N\times M$ STBC matrices
$\mathbf A_\ell$ which are then summed together so that
\begin{eqnarray}\label{eq:LDC}
    \mathbf X=\sum_{\ell =1}^L\mathbf A_\ell s_\ell
\end{eqnarray}
where $L$ is the total number of symbols to be transmitted If $L=
N$, the system is at full-rate (rate-one). At the time slot $n$, the
$n$th row of the coding matrix $\mathbf X$ feeds the $M$ coded
symbols to the $M$ antennas  for transmission. Each of these
transmitter antennas is linked to the receiver antenna through a
channel path coefficient $h_m,~m=1,\cdots,M$. At the receiver of
such a system, for every $N$ time slots ($n=1,\cdots,N$), we receive
an $N$-dimensional signal vector $\mathbf y=[y_1~y_2~\cdots~y_N]^T$
which, as a special case of Eq.~(\ref{eq:MIMOsystem}), can then be
written as
\begin{eqnarray}\label{eq:model}
    \mathbf y={\mathbf X}{\mathbf h}+\boldsymbol \xi
\end{eqnarray}
where ${\mathbf h}=[h_1,\cdots,h_M]^T$ is an $M\times 1$ channel
vector assumed to be circularly symmetric complex Gaussian
distributed with zero-mean and covariance matrix
${\boldsymbol\Sigma}$, and $\boldsymbol \xi$ is an $N\times 1$ noise
vector assumed to be circularly symmetric complex Gaussian with
covariance $\sigma^2\mathbf I_N$. Putting Eq.~(\ref{eq:LDC}) into
Eq.~(\ref{eq:model}), writing the symbols to be transmitted as a
vector and aligning the code-channel products to form the new
channel matrix we can write
\begin{eqnarray}\label{eq:equal_channel}
    \boldsymbol{\mathcal H}=\begin{pmatrix} \mathbf A_1\mathbf
    h&\mathbf A_2\mathbf h&\cdots&\mathbf A_L\mathbf h
    \end{pmatrix}\quad {\rm and}\quad  \mathbf s=[s_1~ s_2~ \cdots
    ~s_L]
\end{eqnarray}
the received signal vector can now be written as
\begin{eqnarray}\label{eq:equal_model}
    \mathbf y=\boldsymbol{\mathcal H}\mathbf s+\boldsymbol \xi
\end{eqnarray}
In this paper, we emphasize on the application of \emph{linear}
receivers for the MISO system in Eq.~(\ref{eq:equal_model}). In the
following, we will derive a condition on the equivalent channel
$\boldsymbol{\mathcal H}$ that renders full-diversity when the
signals are received by a linear receiver. First, we present the
following properties of the
equivalent channel matrix $\boldsymbol{\mathcal H}$: 



\property \label{pro:general-det} Suppose the equivalent channel
$\boldsymbol {\mathcal H}$ in Eq.~\eqref{eq:equal_channel} is such
that $\boldsymbol {\mathcal H}^H\boldsymbol {\mathcal H}$ is
non-singular for any nonzero $\mathbf h$. Then we have the following
inequality:
\begin{eqnarray}\label{eq:bound}
     C_{\min}\|\mathbf h\|^{2L}\le \det\left( \boldsymbol{\mathcal
H}^H\boldsymbol{\mathcal H}\right)\le C_{\max}\|\mathbf h\|^{2L}
\end{eqnarray}
where $C_{\min}$ and $C_{\max}$ are positive constants independent
of $\mathbf h$. \hfill\QED\\
\noindent\proof Since $\mathbf h$ is nonzero, we normalize the
$L\times L$ matrix $\boldsymbol{\mathcal H}^H\boldsymbol{\mathcal
H}$ by dividing each of its elements with $\|\mathbf h\|^2$, i.e., $
\boldsymbol{\mathcal H}^H\boldsymbol{\mathcal H}=\|\mathbf
h\|^2\mathbb H$, where $\mathbb H$ is the normalized matrix with the
$ij$th element being equal to
\begin{eqnarray}\nonumber
    [\mathbb H]_{ij}=\frac{\mathbf h^H}{\|\mathbf h\|}
    \mathbf A_i^H\mathbf A_j
    \frac{\mathbf h}{\|\mathbf h\|}&\qquad i,j=1,2,\cdots,L
\end{eqnarray}
The determinant of positive semi-definite (PSD) matrix $\mathbb H$
is continuous in a closed bounded feasible set $\{\bar{\mathbf
h}:\|\bar{\mathbf h}\|^2=1\}$ where $\bar{\mathbf
h}\triangleq\frac{\mathbf h}{\|\mathbf h\|}$. It has the maximum and
minimum values that are denoted by $C_{\max}$ and $C_{\min}$
respectively. Now, since $\boldsymbol{\mathcal
H}^H\boldsymbol{\mathcal H}$ is non-singular for any nonzero
$\mathbf h$, its determinant is positive. Therefore, $0<C_{\min}\le
C_{\max}$ and Eq.~\eqref{eq:bound} holds. \hfill$\Box$

The following example serves to illustrate the above property.

\textbf{Example 1:} Consider the following channel matrix
    $\boldsymbol{\mathcal H}=\begin{pmatrix}
    h_1&0\\h_2&h_1\\0&h_2
    \end{pmatrix}$.
The determinant of matrix $\boldsymbol{\mathcal
H}^H\boldsymbol{\mathcal H}$ can be written as
\begin{eqnarray}\label{eq:det}
    \det(\boldsymbol{\mathcal H}^H\boldsymbol{\mathcal H})&=&
    \|\mathbf h\|^4\left(
    1-\frac{|h_1|^2}{\|\mathbf h\|^2}
    \frac{|h_2|^2}{\|\mathbf h\|^2}\right)
\end{eqnarray}
Since $\frac{|h_1|^2}{\|\mathbf h\|^2}+
    \frac{|h_2|^2}{\|\mathbf h\|^2}=1$, we can define
    $\frac{|h_1|}{\|\mathbf h\|}=
    \cos\theta$, and $\frac{|h_2|}{\|\mathbf h\|}=\sin\theta$,
and Eq.~\eqref{eq:det} becomes
\begin{eqnarray}
    \det(\boldsymbol{\mathcal H}^H\boldsymbol{\mathcal H})
    &=&
    \|\mathbf h\|^4\left(
    1-\sin^2\theta\cos^2\theta\right)\nonumber\\
    &=&\|\mathbf h\|^4\left(
    1-\frac{1}{4}\sin^2(2\theta)\right)
\end{eqnarray}
It is obvious that the function
$f(\theta)=1-\frac{1}{4}\sin^2(2\theta)$ is continuous in a closed
bounded set. The minimum and maximum of it can be easily obtained as
~$C_{\min}=\frac{3}{4};~~ C_{\max}=1$. Both values are constants and
are independent of the random channel. Thus, the determinant of the
channel matrix is bounded by~
    $\frac{3}{4}\|\mathbf h\|^4\le\det \left(
    \boldsymbol{\mathcal H}^H\boldsymbol{\mathcal H}\right)\le \|\mathbf
    h\|^4$.\hfill$\Box$
\begin{property}\label{pro:inverse-bound}
If $\boldsymbol{\mathcal H}^H\boldsymbol{\mathcal H}$ is
non-singular for any nonzero $\mathbf h$, then the diagonal elements
of $[\boldsymbol{\mathcal H}^H\boldsymbol{\mathcal H}]^{-1}$
satisfies the following inequality
\begin{eqnarray}\label{eq:inverse-bound}
    \left[\left( \boldsymbol{\mathcal H}^H\boldsymbol{\mathcal
    H}\right)^{-1}\right]_{\ell\ell}^{-1}\ge C_0\|\mathbf h\|^{2}
\end{eqnarray}
\end{property}
for $\ell=1,2,\cdots,L$ where $C_0$ is a constant independent of
$\mathbf h$.\hfill\QED\\
 \noindent\proof From the matrix inversion
algorithm~\cite{horn1985}, we have
\begin{eqnarray}\label{eq:inverse}
    \left[\left( \boldsymbol{\mathcal H}^H\boldsymbol{\mathcal
    H}\right)^{-1}\right]_{\ell\ell}^{-1}=
    \frac{\det\left(\boldsymbol{\mathcal H}^H\boldsymbol{\mathcal
    H}\right)}{\det \left(\bar{\boldsymbol{\mathcal H}}_\ell^H\bar{\boldsymbol{\mathcal
    H}}_\ell\right)}
\end{eqnarray}
where $\bar{\boldsymbol{\mathcal H}}_\ell$ is the matrix obtained by
deleting the $\ell$th column vector, $\mathbf A_\ell\mathbf h$, from
$\boldsymbol{\mathcal H}$. We notice that the matrix
$\bar{\boldsymbol{\mathcal H}}_\ell^H\bar{\boldsymbol{\mathcal
    H}}_\ell$ is still PSD and therefore satisfies
the right side inequality of Eq.~\eqref{eq:bound} having an upper
bound denoted by $C_{\ell \max}\|\mathbf h\|^{2(L-1)}$. Applying the
lower bound of Eq.~\eqref{eq:bound} to the numerator and the upper
bound to the denominator of Eq.~(\ref{eq:inverse}), 
we obtain
\begin{eqnarray}\nonumber
    \left[\left( \boldsymbol{\mathcal H}^H\boldsymbol{\mathcal
    H}\right)^{-1}\right]_{\ell\ell}^{-1}&\ge& \frac{C_{\min}\|\mathbf h\|^{2L}}{C_{\ell\max}\|\mathbf
    h\|^{2(L-1)}}\\
    &=&\frac{C_{\min}}{C_{\ell\max}}\|\mathbf h\|^2\ge
    C_0\|\mathbf h\|^2
\end{eqnarray}
where $C_0=C_{\min}/\bar C_{\ell\max}$, with $\bar
C_{\ell\max}=\max\{ C_{\ell\max}, \ell=1,2,\cdots, L\}$.
\hfill$\Box$

\noindent Properties \ref{pro:general-det} and
\ref{pro:inverse-bound} are of fundamental importance to the design
of full diversity STBC for a MISO system employing a linear
detector. This will be presented in the following section.

\section{Diversity Gain of STBC for a MISO System Employing a Linear
Receiver}

Let us first review the concept of diversity gain with reference to
a MIMO system. Consider the MIMO system in Eq.~(\ref{eq:MIMOsystem})
equipped with a maximum likelihood (ML) detector. It is well-known
that an upper bound for the average \emph{pair-wise} error
probability is given by~\cite{tarokh98}
\begin{eqnarray}
    P\left({\mathbf s}\rightarrow{\mathbf s}'\right)
    &\le& \frac{1}{2}~{\rm det}\left({\mathbf I}_{M}
    + \frac{\rho}{8M}{\mathbf X}^H({\mathbf e})
    {\mathbf X}({\mathbf e})\right)^{-{M_R}}\nonumber\\
    &<& \frac{1}{2}\left(\prod_{m=1}^r
    \lambda_m\right)^{-{M_R}}
    \left(\frac{\rho}{8M}\right)^{-r{M_R}}\label{eq:chernoff}
\end{eqnarray}
where $\rho={\rm{E~tr}}[\mathbf X^H \mathbf
X]/{\rm{E~tr}}[\boldsymbol \Xi^H \boldsymbol \Xi]$ is the SNR,
${\mathbf e}={\mathbf s}-{\mathbf s}'$ with ${\mathbf s},{\mathbf
s}'\in \mathcal S$ is the error vector, $r$ $(\le M)$ is the rank,
and $\{\lambda_m\}, m=1, \cdots, r$ are the non-zero eigenvalues of
the matrix $\mathbf X^H({\mathbf e})
\mathbf X({\mathbf e})$. 
The middle part of Eq.~(\ref{eq:chernoff}) is the Chernoff bound,
which at high SNR, can be further tightly bounded by the right side.
For a given $M_R$, two factors dictate the minimization of this
bound on the right side of Eq.~(\ref{eq:chernoff}):
\begin{enumerate}[a)]
    \item The Rank of $\mathbf X^H({\mathbf e})
    \mathbf X({\mathbf e})$: The exponent $r{M_R}$ of the second term
    governs the behaviour of the upper bound with respect to SNR
    and is known as the \emph{diversity gain}. To keep the upper
    bound as low as possible, we should make the diversity gain as large as
    possible. Full diversity is achieved when $r=M$, i.e.,
    $\mathbf X({\mathbf e})$ is of full column rank. This implies
    that
    the diversity gain achieved by an ML detector depends on $\mathbf e$,
    which
    is decided by the type of signalling.
    \item The Determinant of $\mathbf X^H({\mathbf e})
    \mathbf X({\mathbf e})$: The first term consists of the product of the
    non-zero eigenvalues of $\mathbf X^H({\mathbf e})
    \mathbf X({\mathbf e})$ and is called the
    \emph{coding gain}. For $\mathbf X^H({\mathbf e})
    \mathbf X({\mathbf e})$ being full rank, this product is its determinant
    the minimum value of which (taken over \emph{all}
    distinct symbol vector pairs $\{{\mathbf s}, {\mathbf s}'\}$) must be maximized.

\end{enumerate}
At high SNR, the upper bound in Eq.~(\ref{eq:chernoff}) is dominated
by the exponent $-rM_R$ of $\rho$. This leads to a more general
definition of diversity gain~\cite{zheng03} as being \emph{the total
degrees of freedom offered by a communication system, reflected by
the factor involving the negative power of the SNR in the expression
of the error probability.}  Full diversity gain is achieved when the
total degrees of freedom ($=MM_R$) offered in the multi-antenna
system are utilized. We adopt this latter notion of diversity gain
when we consider the STBC for the MISO system.

Since $M_R=1$, full diversity for a MISO system is achieved if the
exponent of the SNR in the expression of the error probability is
equal to $-M$. Let us now consider the condition on
$\boldsymbol{\mathcal H}$ for which full-diversity is achieved by a
MISO system employing a \emph{linear} receiver. We need only to
consider the use of a linear zero-forcing (ZF) receiver because the
same condition extends to MISO systems using linear minimum mean
square (MMSE) receivers or other more sophisticated receivers.
Since the diversity gain of a communication system relates the
probability of error to SNR, we first analyze the symbol error
probability (SEP) of detecting different signal constellations by a
linear ZF equalizer and express these in terms of the SNR.

\subsection{Symbol Error Probability of Various Signalling Schemes}
\noindent Here, we examine three commonly used signalling schemes:
1) square QAM, 2) PAM and 3) PSK constellations respectively. Let
$\mu$ denote the cardinality. Firstly, we summarize the definition
of some common parameters which govern the performance of the ZF
linear detectors under these schemes. We use the index $i=1,~2,~3$
to denote parameters associated with the three signalling schemes as
ordered above. Let $E_{si},~i=1,~2,~3$, denote the respective
average symbol energy in each of the above schemes, and let
$\sigma^2$ be the noise variance at the receiver antenna. Therefore,
the SNR for each symbol at the receiver is given by
\begin{equation}\label{eq:symbolSNR}\vspace{-.3cm}
    \rho_i = E_{si}/\sigma^2; ~~\qquad i = 1,~2,~3
\end{equation}
\noindent 
Note that $\sigma^2[\boldsymbol{\mathcal H}^H\boldsymbol{\mathcal
H}]^{-1}_{\ell\ell}$ is the noise power at the output of the ZF
equalizer for the $\ell$th symbol.

\begin{enumerate}[1.]
\item\emph{Square QAM signals}: The SEP of a ZF receiver for the
square QAM signal $s_\ell$ is~\cite{simon98}
\begin{eqnarray}\label{eq:qamsep}\small
  &&{P}_{\rm 1}({\mathbf h},s_\ell)=\nonumber\\&&\hspace{-.3cm}4\left(1-
  \frac{1}{\sqrt{\mu}}\right)Q\left(\sqrt{\frac{3
  E_{s1}}{2(\mu-1)\sigma^2\left[\left(\boldsymbol{\mathcal H}^H\boldsymbol{\mathcal H}\right)^{-1}
  \right]_{\ell\ell}}}\right)\nonumber\\
  \hspace{-.2cm}&-&\hspace{-.3cm}4\hspace{-.1cm}\left(1-\frac{1}{\sqrt{\mu}}\right)^2\hspace{-.2cm}Q^2\hspace{-.2cm}
  \left(\sqrt{\frac{3E_{s1}}
  {2(\mu-1)\sigma^2\left[\left(\boldsymbol{\mathcal H}^H\boldsymbol{\mathcal H}\right)^{-1}
  \right]_{\ell\ell}}}\right)\hspace{.3cm}~~
\end{eqnarray}
where 
$Q(z)=\frac{1}{\sqrt{2\pi}}\int_{z}^\infty e^{-x^2/2}dx$. We use the
following alternative expressions for the $Q$- and
$Q^2$-functions~\cite{craig91,pawula82,weinstein74,simon98}
\begin{eqnarray}\label{eq:q}
    Q(z)&=&\frac{1}{\pi}\int_0^{\pi/2}\hspace{-.2cm}
    \text{exp}\left(-\frac{z^2}{2\sin^2\theta}\right)\,d\theta \quad z\ge 0\\
    \label{eq:qq}
    Q^2(z)&=&\frac{1}{\pi}\int_0^{\pi/4}\hspace{-.2cm}
    \text{exp}\left(-\frac{z^2}{2\sin^2\theta}\right)\,d\theta \quad z\ge 0
\end{eqnarray}
\noindent Substituting Eqs.~(\ref{eq:q}) and~(\ref{eq:qq}) into
Eq.~(\ref{eq:qamsep}) and after a little manipulation, we obtain
\begin{eqnarray}\label{eq:aveqamsep}
   &&{P}_{\rm 1}({\mathbf h}, s_\ell)
    =\frac{4}{\pi}\left(1-\frac{1}{\sqrt{\mu}}\right)\nonumber\\
    &&\int_0^{\pi/4}\hspace{-.4cm}\exp\left(-\frac{3E_{s1}}
    {4(\mu-1)\sigma^2
    \left[\left(\boldsymbol{\mathcal H}^H\boldsymbol{\mathcal H}\right)^{-1}
  \right]_{\ell\ell}\sin^2\theta}\right)d\theta\nonumber\\
   \hspace{-.2cm}
   &+&\frac{4}{\pi \sqrt{\mu}}\left(1-\frac{1}{\sqrt{\mu}}\right)\nonumber\\
  \hspace{-.4cm} &&\hspace{-.4cm}
    \int_{\pi/4}^{\pi/2}\hspace{-.4cm} \exp\hspace{-.1cm}\left(\hspace{-.1cm}-\frac{3E_{s1}}{4(\mu-1)
    \sigma^2\left[\left(\boldsymbol{\mathcal H}^H\boldsymbol{\mathcal H}\right)^{-1}
  \right]_{\ell\ell}\sin^2\theta}\right)\hspace{-.1cm}d\theta
\end{eqnarray}
We can obtain an upper bound for Eq.~(\ref{eq:aveqamsep}) by putting
$\sin \theta = 1$ in the two integrals and, writing $\rho
_1=E_{s1}/\sigma^2$, this easily simplifies to
\begin{eqnarray}\label{eq:qamsep-UB}
    {P}_{\rm 1}({\mathbf h}, s_\ell)\le \frac{\mu-1}
    {\mu}\exp\hspace{-.1cm}\left(\hspace{-.1cm}-\frac{3\rho_1}{4(\mu-1)\left
    [\left(\boldsymbol{\mathcal H}^H\boldsymbol{\mathcal H}\right)^{-1}
  \right]_{\ell\ell}}\right)
\end{eqnarray}

\item\emph{PAM signals}: The SEP of the ZF receiver for a
$\mu$-ary PAM signal $s_\ell$ is given by~\cite{simon98}
\begin{equation}\label{eq:pamsep}
    {P}_{\rm 2}({\mathbf h}, s_\ell)=\frac{2(\mu -1)}
    {\mu}Q\left(\sqrt{\frac{3E_{s2} }{(\mu^2-1)\sigma^2
    \left[\left(\boldsymbol{\mathcal H}^H\boldsymbol{\mathcal H}\right)^{-1}
  \right]_{\ell\ell}}}\,\right)
\end{equation}
Now, using Eq.~(\ref{eq:q}) and noting that by putting $\sin \theta
= 1$ in the integral, we have $Q(z)\le {\rm exp}(-z^2)$ for $z\ge
0$. Together with $\rho_2=E_{s2}/\sigma^2$, we arrive at an upper
bound of Eq.~(\ref{eq:pamsep}),
\begin{equation}\label{eq:pamsep-UB}
    {P}_{\rm 2}({\mathbf h}, s_\ell)\le \frac{\mu-1}
    {\mu}\exp\left(-\frac{3\rho_2} {2(\mu^2-1)\left[
    \left(\boldsymbol{\mathcal H}^H\boldsymbol{\mathcal H}\right)^{-1}
  \right]_{\ell\ell}}\right)
\end{equation}

\item\emph{PSK signals}: The SEP of a ZF receiver for the PSK
signal $s_{k\ell}$ is given by~\cite{simon98}
\begin{equation}\label{eq:psksep}
    P_{\rm 3}({\mathbf h},
    s_\ell)=\frac{1}{\pi}\int_0^{(\mu-1)\pi/\mu}\hspace{-.6cm}
    \text{exp}\hspace{-.1cm}\left(\hspace{-.1cm}-\frac{ E_{s3}\sin^2(\pi/\mu)}{2\sigma^2\left
    [\left(\boldsymbol{\mathcal H}^H\boldsymbol{\mathcal H}\hspace{-.1cm}\right)^{-1}
  \right]_{\ell\ell}\sin^2\theta}\right)\hspace{-.15cm}d\theta
\end{equation}
Writing $\rho_2=E_{s2}/\sigma^2$, similar to the PAM signal,
Eq.~(\ref{eq:psksep}) can be upper bounded by
\begin{equation}\label{eq:psksep-UB}
    P_{\rm 3}({\mathbf h}, s_\ell)\le\frac{(\mu-1)}
    {\mu}\text{exp}\left(-\frac{\rho_3\sin^2(\pi/\mu)}
    {2\left[\left(\boldsymbol{\mathcal H}^H\boldsymbol{\mathcal H}\right)^{-1}
  \right]_{\ell\ell}}\right)
\end{equation}
\end{enumerate}

\subsection{Design Criterion for Full-Diversity STBC for
a MISO System with Linear Receivers\label{sec:full-diversity}} We
now examine the \emph{diversity gain} achievable by a MISO system.

%
\begin{theorem}\label{theo:full-diversity}
    For a MISO system employing a square QAM, a PAM, or a PSK
    signalling scheme of cardinality $\mu$ in the transmission, a
    linear receiver (ZF/MMSE) achieves \emph{full diversity} for the system
    if $\boldsymbol{\mathcal H}^H\boldsymbol{\mathcal H}$ is non-singular
    for any nonzero $\mathbf h$, or equivalently, if
    $\mathbf X^H(\mathbf s)\mathbf X(\mathbf s)$ is non-singular
    for any nonzero $\mathbf s$.\hfill\QED
\end{theorem}
\textit{Proof}: From Eqs.~(\ref{eq:qamsep-UB}),
(\ref{eq:pamsep-UB}), and (\ref{eq:psksep-UB}), we can arrive at a
generalized upper bound on the symbol error probability for the
${\mu}$-ary QAM, PAM, and PSK signals such that\vspace{-.2cm}
\begin{multline}\label{eq:unified sep}
    {P}_i({\mathbf h}, s_\ell)\le \frac{\mu-1}{\mu}
    \exp\left(-\frac{a_i \rho_i}
    {\left[\left(\boldsymbol{\mathcal H}^H\boldsymbol{\mathcal H}\right)^{-1}
    \right]_{\ell\ell}}\right),\\ i=1,~2,~3
\end{multline}
where $\rho _i=E_{si}/\sigma^2$, and\vspace{-.2cm}
\begin{equation}\label{eq:a_i}
    a_{\rm 1}=3/[4(\mu-1)],a_{\rm 2}=3/[2(\mu^2-1)],
    {\rm and}~
    a_{\rm 3} =\sin^2(\pi/\mu)/2
\end{equation}
Since $\boldsymbol{\mathcal H}^H\boldsymbol{\mathcal H}$ is
non-singular for any nonzero $\mathbf h$, we can apply
Eq.~\eqref{eq:inverse-bound} on Eq.~(\ref{eq:unified sep}). Here, we
see that the arithmetic mean of the SEP of all the three signalling
schemes have a general upper bound given by
\begin{eqnarray}
    {P}_i({\mathbf h})&\le& \frac{\mu-1}{\mu}
    \exp\left(-a_i\rho_i C_0\|{\mathbf h}\|^2\right)\nonumber\\
    &=&\frac{\mu-1}{\mu}
    \exp\left(-a_i\rho_i C_0{\mathbf h}^H {\mathbf h}\right)\label{eq:unified bound}
\end{eqnarray}
Now, $\mathbf h$ is assumed to be Gaussian with zero mean and
covariance matrix $\boldsymbol\Sigma$. Therefore, averaging the
exponential part of the right side of Eq.~(\ref{eq:unified bound})
over the density function of ${\mathbf h}$ yields
\begin{eqnarray}\label{eq:avUB}
    \frac{1}{\pi^M\rm{det}\boldsymbol \Sigma}
    \int \exp\left(-a_i\rho_i C_0{\mathbf h}^H {\mathbf h}\right)
    \rm{exp}\left(-\mathbf h^H\boldsymbol\Sigma^{-1}\mathbf h \right)
    d\mathbf h 
    \nonumber\\
    =
    \left[\frac{{\rm det}\left((a_i\rho_i C_0\mathbf
    I+\boldsymbol\Sigma^{-1})^{-1}\right)}
    {\rm{det}\boldsymbol\Sigma}\right]
    =\det({\mathbf I}+a_i\rho_i C_0\boldsymbol\Sigma)^{-1}
\end{eqnarray}
Substituting Eq.~(\ref{eq:avUB}) into Eq.~(\ref{eq:unified bound}),
we establish the following inequalities:
\begin{eqnarray}\label{eq:average unification sep}
    {\rm E}\left[{P}_i({\mathbf h})\right]&\le&\frac{\mu-1}
    {\mu}\det({\mathbf I}+a_i\rho_i C_0\boldsymbol\Sigma)^{-1}\nonumber\\
    &\le&
    \left (\frac{\mu-1}{\mu}\det(C_0\boldsymbol\Sigma)^{-1}{a_i}^{-M}\right)
    {\rho_i}^{-M},\nonumber\\
    &&~\hspace{2cm} i=1,2,3
\end{eqnarray}
The exponent of $\rho_i$ in Eq.~(\ref{eq:average unification sep})
indicates that the upper bound of the SEP using a ZF receiver in a
MISO system to detect signals from the three schemes indeed achieves
full diversity for non-singular $\boldsymbol{\mathcal
H}^H\boldsymbol{\mathcal H}$.

\noindent We now show the equivalency of the following two
statements:
\begin{enumerate}[1)]
\item $\boldsymbol{\mathcal H}^H\boldsymbol{\mathcal H}$ is
non-singular for any nonzero $\mathbf h$; and

\item $\mathbf X^H(\mathbf s)\mathbf X(\mathbf s)$ is
non-singular for any nonzero $\mathbf s$.
\end{enumerate}
We will show 1) $\Rightarrow$ 2), and the reverse can be similarly
proved. From the development of Eq.~\eqref{eq:equal_model}, we have
$\mathbf X(\mathbf s)\mathbf h=\boldsymbol{\mathcal H}\mathbf s$.
Now, if $\boldsymbol{\mathcal H}^H\boldsymbol{\mathcal H}$ is
non-singular for any nonzero $\mathbf h$, then $\boldsymbol{\mathcal
H}$ has full column rank, and hence $\boldsymbol{\mathcal H}\mathbf
s\ne \mathbf 0~ \forall~ \mathbf s\ne 0$. Therefore $\mathbf
X(\mathbf s)\mathbf h\ne \mathbf 0$ for any $\mathbf s\ne 0$,
$\mathbf h\ne 0$. This implies full column rank of matrix $\mathbf
X(\mathbf s)$, and hence $\mathbf X^H(\mathbf s)\mathbf X(\mathbf
s)$ is non-singular for any nonzero $\mathbf s$.\hfill$\Box$

\noindent\underline{Remarks on Theorem 1}:
\begin{enumerate}[a)]
\item Although the proof provided here is for square QAM, PAM and
PSK signallings, Theorem~\ref{theo:full-diversity} can be shown to
be valid for any signal constellation.

\item Since the condition provided here is sufficient for a linear
receiver to achieve full diversity, the same condition naturally
yields full diversity for more sophisticated receivers such as
MMSE/ZF-DFE or ML receivers.

\item For a MIMO system using an ML detector, the
requirement for full diversity as indicated by
Eq.~(\ref{eq:chernoff}) is that the coding matrix $\mathbf
X^H({\mathbf e})\mathbf X({\mathbf e})$ is maintained at full-rank
for the signals $\mathbf s, \mathbf s'\in\mathcal S$. However, for a
MISO system employing a linear receiver,
Theorem~\ref{theo:full-diversity} shows us that full diversity is
achieved if the coding matrix $\mathbf X^H({\mathbf s})\mathbf
X({\mathbf s})$ is of full rank for \emph{any} signal $\mathbf s$, a
much stronger condition than that required by systems using an ML
detector.
\end{enumerate}

\noindent In the following section, we present the \emph{Toeplitz
STBC} which has a simple structure satisfying the full-rank
condition in Theorem~\ref{theo:full-diversity} and is therefore a
\emph{full-diversity} STBC for a MISO system employing a linear
receiver.

\section{Toeplitz Space-Time Block Codes and Their Properties}\label{sec:ToeplitzSTBC}
\subsection{Toeplitz STBC for a MISO System~\cite{zlw-isit05}}

To examine the structure of the Toeplitz space-time block code, we
let ${\boldsymbol \alpha}=[\alpha_1~~\alpha_2~~ \cdots~~
\alpha_L]^T$. A $(K+L-1)\times K$ Toeplitz matrix generated by
${\boldsymbol \alpha}$ and a positive integer $K$, denoted by
${\mathcal T}({\boldsymbol \alpha}, L, K)$, is defined as
\begin{eqnarray}\label{eq:toeplitz}
    [{\mathcal T}({\boldsymbol \alpha},L,K)]_{i
    j}=\left\{\begin{array}{cc}
    \alpha_{i-j+1}, & \textrm{if $i\ge j$ and $i-j<L$} \\
    0, & \textrm{otherwise}\\
\end{array} \right.
\end{eqnarray}
which can be explicitly written as
 \begin{eqnarray}\label{eq:toep}
    {\mathcal T}({\boldsymbol \alpha},L,K)=\left(
    \begin{array}{cccc}
        \alpha_1 & 0 & \ldots & 0\\
        \alpha_2 &\alpha_1 & \ldots & 0\\
        \vdots &\alpha_2 & \ddots & \vdots\\
        \alpha_L & \ddots & \ddots &\alpha_1\\
        0 & \ddots & \ddots &\alpha_2\\
        \vdots & \ddots & \ddots & \vdots\\
        0 & \ddots & 0 &\alpha_L
    \end{array}\right)_{(K+L-1)\times K}
\end{eqnarray}
If we replace $\boldsymbol \alpha$ by $\mathbf s$, the information
symbols to be transmitted, then a \emph{Toeplitz} STBC matrix
${\mathcal X}_{\mathbf B}({\mathbf s})$ is defined as
\begin{eqnarray}\label{eq:toeplitz coding matrix}
    {\mathcal X}_{\mathbf B}({\mathbf s})={\mathcal T}({\mathbf s},L,
    K)\cdot{\mathbf B}
\end{eqnarray}
where, for $K \le M$, ${\mathbf B}$ is a $K\times M$ matrix of rank
$K$ placed in the coding matrix to facilitate the transmitter
antennas with beamforming capability. At time slot $n$, the $n$th
row of the $N\times M$ matrix ${\mathcal X}_{\mathbf B}({\mathbf
s})$ is fed to the $M$ transmitter antennas for transmission. Apply
the Toeplitz space-time block coding
matrix to the MISO system described in Eq.~\eqref{eq:model}, 
and we have
\begin{equation}\label{eq:model2}
    {\mathbf y}= {\mathcal T}(\widetilde{\mathbf h},K,L)\,{\mathbf
    s}+{\boldsymbol\xi}
\end{equation}
where $\widetilde{\mathbf h}={\mathbf B}{\mathbf h}$, and $L=N-K+1$.
Thus, ${\mathcal T}(\widetilde{\mathbf h},K,L)$ can be viewed as the
overall channel matrix of the MISO system.

{\bf{Example 2}}. For $K=M=L=2, N=K+L-1=3$, and ${\mathbf
B}={\mathbf I}_2$, the codeword matrix and channel matrix are,
respectively,
\begin{eqnarray}\nonumber
    {\mathcal X}_{{\mathbf I}_2}({\mathbf s})=\left(%
    \begin{array}{cc}
        s_1 & 0   \\
        s_2 & s_1 \\
        0   & s_2 \\
    \end{array}%
    \right), & {\mathcal T}(\widetilde{\mathbf h},2,
    2)=\left(%
    \begin{array}{cc}
        h_1 & 0   \\
        h_2 & h_1 \\
        0   & h_2 \\
    \end{array}%
    \right)
\end{eqnarray}
For this code, there are $L=2$ symbols to be transmitted in $N=3$
channel uses. Therefore, the symbol transmission rate of this
system is $R_s=\frac{2}{3}$ symbols per channel use.

\noindent\underline{Remarks on Toeplitz STBC}:
\begin{enumerate}[a)]
    \item Eq.~(\ref{eq:model2}) is identical in form to that
    describing a MIMO intersymbol interference channel for zero-padding
    block data transmission~(e.g.~\cite{scaglione99}). It can thus be
    interpreted that the original MISO channel is transformed into a
    Toeplitz virtual MIMO channel. In other words,
    the space diversity has been exchanged for delay (time) diversity. This is realized
    by transforming the flat fading channel into a frequency selective channel with
    zero-padding. This technique is parallel to that employed
    in~\cite{winter94}.
    \footnote{In \cite{winter94}, consideration is given only to the use of an ML detector
    and performance analysis is restricted to BPSK signalling under the assumption that
    only one bit error occurs.

}

    \item For such a system, we can utilize the
    efficient Viterbi algorithm~\cite{viterbi67} to detect the signal
    ${\mathbf s}$ if perfect channel knowledge is available at the
    receiver. On the other hand, when channel
    coefficients are not known at the receiver, we can make use of the
    second order statistics of the received signal to blindly identify
    the channel~\cite{tong98,scaglione99}.


    \item Toeplitz STBC is a \emph{non-orthogonal} STBC whose coding matrix
    $\mathbf X^H\mathbf X$ possesses non-vanishing determinant for
    \emph{any} signalling scheme. Hence according to Theorem~\ref{theo:full-diversity},
    the code achieves full diversity even with the use of a linear receiver.
    On the other hand, since full diversity STBC designed for ML receivers (e.g.,
    \cite{quasi-dao,quasi-sharma,quasi-xia,G-A-code}) maintain non-vanishing
    determinant only for certain types of signalling, full diversity gain
    is not guaranteed when a linear receiver is used.

    \item When $\mathbf B=\mathbf I$, Toeplitz STBC becomes a special delay
    diversity code (DDC)~\cite{winter-DCC, Paulraj-DCC} with padded zeroes.
    In general, DDC is applied with the use of outer channel coding and
    ML detectors to achieve the full diversity gain. However, here
    we show that the Toeplitz STBC possesses special properties which enable full space
    diversity to be achieved even with the use of the simplest linear receiver and
    the signals can be of any type.

\end{enumerate}\vspace{-.3cm}

\subsection{Properties of Toeplitz STBC}\label{sec3:properties}
We now examine some important properties of the Toeplitz space-time
block codes introduced in the previous subsection. These properties
will be useful in performance analysis and code designs in the
ensuing sections.

\begin{property}\label{pro:full-rate}
The definition of the Toeplitz space-time code shows that the symbol
transmission rate is $R_s=\frac{L}{N}=\frac{N-K+1}{N}$ symbols per
channel use when $K\le M$. Therefore, for a fixed $M$, the
transmission rate $R$ can approach unity if the number of channel
uses is sufficiently large.
\end{property}

\begin{property}\label{pro:toeplitz lower bound}
For any nonzero vector ${\boldsymbol \alpha}$, there exists
$0<C_{T\min}\le C_{T\max}\le 1$, and the matrix $\left({\mathcal
T}^H({\boldsymbol \alpha},L,K){\mathcal T}({\boldsymbol
\alpha},L,K)\right)$ satisfies the following inequality,
\begin{eqnarray}
    C_{T\min}\|{\boldsymbol \alpha}\|^{2K}&\le& \det\left({\mathcal
    T}^H({\boldsymbol \alpha},L,K){\mathcal T}({\boldsymbol
    \alpha},L,K)\right)\nonumber\\
    &\le& C_{T\max}\|{\boldsymbol \alpha}\|^{2K}\label{eq:toeplitz inequality}
\end{eqnarray}

\end{property}
\textit{Proof}: By letting $\boldsymbol \alpha=\mathbf h$ in
Eq.~(\ref{eq:equal_channel}) and choosing
\begin{eqnarray}\nonumber
    \mathbf A_\ell=\mathbf P^{\ell-1} \mathbf A_0,\qquad \ell=1,\cdots, L
\end{eqnarray}
where
\begin{eqnarray}\nonumber
    \mathbf P=\begin{pmatrix}
    \mathbf 0_{(L-1)\times 1}& 1\\
    \mathbf I_{(L-1)}&\mathbf 0_{1\times(L-1)}
    \end{pmatrix}\qquad
    \mathbf A_0=\begin{pmatrix}
    \mathbf I_L\\ \mathbf 0_{(K-1)\times L}
    \end{pmatrix}
\end{eqnarray}
we obtain an equivalent channel $\boldsymbol {\mathcal H}$ of the
same structure as ${\mathcal T}({\boldsymbol \alpha},L,K)$. Hence,
${\mathcal T}({\boldsymbol \alpha},L,K)$ is a special case of
$\boldsymbol {\mathcal H}$. Thus, from Property
\ref{pro:general-det}, there exist $C_{T\min}$ and $C_{T\max}$ for
which Eq.~\eqref{eq:toeplitz inequality} holds. Now, we note that
the diagonal entries of the matrix ${\mathcal T}^H({\boldsymbol
\alpha},L,K){\mathcal T}({\boldsymbol \alpha},L,K)$ are all the same
and are equal to
    $\left[{\mathcal T}^H({\boldsymbol\alpha},L,K){\mathcal
    T}({\boldsymbol\alpha},L,K)\right]_{kk}=\|{\boldsymbol
    \alpha}\|^2,~ k=1,\cdots,K$.
By applying Hadamard's inequality~\cite{horn1985}, we arrive at:
\begin{eqnarray}&&\det\left({\mathcal
    T}^H({\boldsymbol \alpha},L,K){\mathcal T}({\boldsymbol
    \alpha},L,K)\right)\le \|{\boldsymbol
    \alpha}\|^{2K}\qquad\nonumber\\
    &&\Rightarrow~~C_{T\max}\le 1\label{eq:upper_det}
\end{eqnarray}


\noindent Furthermore, since $\boldsymbol\alpha$ is nonzero, we can
assume, without loss of generality, that the first element
$\alpha_1\neq 0$. (Otherwise, we can always permute the nonzero
element to the first position.) The $N\times K$ ``tall" matrix,
${\mathcal T}({\boldsymbol\alpha},L,K)$, can be partitioned into a
top $K\times K$ matrix $\boldsymbol\Omega_1$, and a bottom matrix
$\boldsymbol\Omega_2$ containing the rest of ${\mathcal
T}({{\boldsymbol\alpha}},L,K)$, i.e.,
\begin{eqnarray}\label{eq:partitionT}
    {\mathcal T}({{\boldsymbol\alpha}},L,K)=
    {\left(\begin{array}{cccc}
    \alpha_1 & 0 & \ldots & 0\\
    \alpha_2 & \alpha_1 & \ddots & \vdots\\
    \vdots & \alpha_2 & \ddots & 0\\
    \alpha_K & \alpha_{K-1} & \cdots & \alpha_1\\
    --&--&--&--\\
    \alpha_{K+1} & \alpha_K & \cdots & \alpha_2\\
    \vdots & \vdots & \cdots & \vdots\\
    \alpha_L & \alpha_{L-1} & \cdots & \alpha_{L-K-1}\\
    0 & \alpha_L & \cdots & \alpha_{L-K}\\
    \vdots & \ddots & \ddots & \vdots\\
    0 & \cdots & 0 & \alpha_L
    \end{array}\right)}
    \begin{array}{c}
    \boldsymbol\Omega_1\\\\\\\\\\
    \boldsymbol\Omega_2
    \end{array}\nonumber
\end{eqnarray}
We note that $ \boldsymbol\Omega_1$ is a lower triangular matrix
having equal diagonal elements $\alpha_1\neq 0$. (Here, we assume
that $K\le L$. The proof is equally valid if $L\le K$ by exchanging
the roles of $K$ and $L$). Since ${\mathcal
T}^H({{\boldsymbol\alpha}},L,K){\mathcal
T}({{\boldsymbol\alpha}},L,K) =
{\boldsymbol\Omega_1}^H{\boldsymbol\Omega_1}+{\boldsymbol\Omega_2}^H{\boldsymbol\Omega_2}$,
using a standard result \cite[p.~484]{horn1985} on the determinant
of the sum of a positive definite and a positive semi-definite
matrix we have
\begin{eqnarray}
    &&{\rm det}\left({\mathcal T}^H({{\boldsymbol\alpha}},L,K){\mathcal T}
    ({{\boldsymbol\alpha}},L,K)\right)\nonumber\\&&\ge
    {\rm det}({\boldsymbol\Omega_1}^H{\boldsymbol\Omega_1})+{\rm det}
    ({\boldsymbol\Omega_2}^H{\boldsymbol\Omega_2})
    \nonumber\\&&\ge\det({\boldsymbol\Omega_1}^H{\boldsymbol\Omega_1})\nonumber\\
    &&=|\alpha_1|^{2K}\label{eq:detTgamma}
\end{eqnarray}
Since $\alpha_1\neq 0$, $\det\left({\mathcal T}^H({\boldsymbol
\alpha},L,K){\mathcal T} ({\boldsymbol \alpha},L,K)\right)$ is
nonzero for any nonzero $\boldsymbol\alpha$ and thus we have
\begin{equation}\label{eq:CTmin}
    C_{T\min}> 0
\end{equation}
The conclusions in Eqs.~(\ref{eq:upper_det}) and (\ref{eq:CTmin})
complete the proof of the property.\hfill$\Box$

\noindent We see that $\mathcal T$ is non-singular by
Property~\ref{pro:toeplitz lower bound}, then using the fact that
$\mathcal T$ as a special case of $\mathcal H$ in
Property~\ref{pro:inverse-bound}, we have:
\begin{property}\label{pro:zf-lemma}
 \begin{multline}\label{eq:invers-bound-teo}
    \big[\big({\mathcal T}^H({\boldsymbol\alpha},L,K){\mathcal T}
    ({\boldsymbol\alpha},L,K)\big)^{-1}\big]_{kk}^{-1} \ge
    C_{T\min}\|{\boldsymbol\alpha}\|^2\\{\rm for}~~ k=1, 2,
\cdots, K
\end{multline}
where the existence of $C_{T\min}>0$ is shown in
Property~\ref{pro:toeplitz lower bound}, holds.~\hfill\QED
\end{property}

We now introduce the following definition related to the measure in
a signal constellation $\mathcal S$:
\begin{definition}
    For $\mathbf s, \mathbf s'\in {\mathcal S}$, the minimum distance
    of the signal constellation is defined as
    \begin{equation}\label{eq:dmin}
        d_{\rm min}({\mathcal S})=\min\limits_{\mathbf s \ne \mathbf s'}
        \|\mathbf s - \mathbf s'\|
    \end{equation}
\end{definition}
If $\|\mathbf s - \mathbf s'\|=d_{\rm min}({\mathcal S})$, we say
that $\mathbf s$ and $\mathbf s'$ are \emph{neighbours}.

From the definition of the coding matrix ${\mathcal X}$ in
Eq.~(\ref{eq:toeplitz coding matrix}) with $\mathbf B = \mathbf
I_M$, we can now establish a lower bound for a metric between
${\mathcal X}_{{\mathbf I}_M}({\mathbf s})$ and ${\mathcal
X}_{{\mathbf I}_M}({\mathbf s}')$ in $d_{\rm min}({\mathcal S})$.
Let $\mathbf e = (\mathbf s - \mathbf s')$. For notational
convenience, we let ${\mathcal X}_{{\mathbf I}_M}({\mathbf
e},\{i_1,i_2,\cdots,i_m\})$ denote the matrix consisting of $m$ of
the columns of ${\mathcal X}_{{\mathbf I}_M}({\mathbf e})$ indexed
by $\{i_1, i_2, \cdots, i_m\}$ where $i_1<i_2<,\cdots,<i_m$ and
these columns are not necessarily consecutively chosen. Then, we
have
\begin{property}\label{pro:submatrix}
    For $\mathbf s \neq \mathbf s'\in{\mathcal S}^L$, where ${\mathcal
    S}^L= {\mathcal S}\times {\mathcal
S}\cdots\times{\mathcal S}$, and any nonzero vector ${\mathbf e}=
    (\mathbf s - \mathbf s')$, we have
    \begin{multline}\label{eq:detsubmatrix}
        \det\big[{\mathcal X}^H_{{\mathbf I}_M}({\mathbf
        e},\{i_1,i_2,\cdots,i_m\}){\mathcal X}_{{\mathbf I}_M}({\mathbf
        e},\{i_1,i_2,\cdots, i_m\})\big]\\\ge d^{2m}_{\rm min}({\mathcal S})
    \end{multline}
    for $m=0,1,\cdots,M-1$, where the equality holds if and only if
    ${\mathbf s}$ and ${\mathbf s}'$ are neighbours, i.e., iff $\|{\mathbf e}\|
    =d_{\rm min}({\mathcal S})$.~\hfill\QED
\end{property}
\textit{Proof}: The proof of this property is similar to that of
Property~\ref{pro:toeplitz lower bound}. As in
Property~\ref{pro:toeplitz lower bound}, without loss of generality,
we can always assume that $e_1\ne 0$ with $e_1$ being the first
element of $\mathbf{e}$.
Thus, ${\mathcal X}_{{\mathbf I}_M}({\mathbf e})$ can be written as
\begin{equation}\label{eq:X_I_partitioned}
    {\mathcal X}_{{\mathbf I}_M}({\mathbf e})=\left(
    \begin{array}{cccc}
        e_1 & 0 & \ldots & 0\\
        e_2 & e_1 & \ldots & 0\\
        \vdots & e_2 & \ddots & \vdots\\
        e_M & \ddots & \ddots & e_1\\
        \vdots & e_M & \ddots & e_2\\
        e_L & \ddots & \ddots & \vdots\\
        0 & \ddots & \ddots & e_M\\
        \vdots & \ddots & \ddots & \vdots\\
        0 & \ddots & 0 & e_L
    \end{array}\right)_{N\times M}.
\end{equation}
An important observation in Eq.~(\ref{eq:X_I_partitioned}) is that
the top submatrix consisting of the first $M$ rows of ${\mathcal
X}_{{\mathbf I}_M}({\mathbf e})$ is a $M\times M$ lower triangular
matrix with nonzero diagonal entries and therefore, nonzero
determinant. We can also see that the submatrix ${\mathcal
X}_{{\mathbf I}_M}({\mathbf e},\{i_1,i_2,\cdots,i_m\})$ preserves
the same property
because by permuting its rows and columns, an $m\times m$ lower
triangular matrix can always be formed as its top part, i.e.,
${\mathcal X}_{{\mathbf I}_M}({\mathbf e},\{i_1,i_2,\cdots,i_m\})$
can be expressed as
\begin{equation}
    \nonumber {\boldsymbol \Pi}_1{\mathcal X}_{{\mathbf I}_M}({\mathbf
    e},\{i_1,i_2,\cdots,i_m\}){\boldsymbol \Pi}_2=\left(%
    \begin{array}{c}
        {\boldsymbol\Omega_1}\vspace{-.4cm} \\
        --\vspace{-.4cm}\\
        {\boldsymbol\Omega_2} \\
    \end{array}%
    \right)
\end{equation}
where ${\boldsymbol \Pi}_1$ and ${\boldsymbol \Pi}_2$ denote the
$N\times N$ and $m\times m$ permutation matrices, respectively,
$\boldsymbol\Omega_1$ contains the first $m$ rows of ${\mathcal
X}_{{\mathbf I}_M}({\mathbf e},\{i_1,i_2,\cdots,i_m\})$ and hence is
lower triangular, and $\boldsymbol\Omega_2$ denotes the remaining
submatrix of ${\mathcal X}_{{\mathbf I}_M}({\mathbf
e},\{i_1,i_2,\cdots,i_m\})$. Since the permutation of the rows and
columns of ${\mathcal X}_{{\mathbf I}_M}({\mathbf
e},\{i_1,i_2,\cdots,i_m\})$ does not change the determinant of its
autocorrelation matrix ${\mathcal X}_{{\mathbf I}_M}^H({\mathbf
e},\{i_1,i_2,\cdots,i_m\}){\mathcal X}_{{\mathbf I}_M}({\mathbf
e},\{i_1,i_2,\cdots,i_m\})$, therefore, as in
Property~\ref{pro:toeplitz lower bound},
we arrive at
\begin{multline}
    {\rm det}\left[{\mathcal X}_{{\mathbf I}_M}^H({\mathbf
    e},\{i_1,i_2,\cdots,i_m\}){\mathcal X}_{{\mathbf I}_M}({\mathbf
    e},\{i_1,i_2,\cdots,i_m\})\right]\\\ge {\rm det}({\boldsymbol\Omega_1}^H
    {\boldsymbol\Omega_1})+{\rm det}({\boldsymbol\Omega_2}^H{\boldsymbol\Omega_2})\ge
    d^{2n}_{\rm min}({\mathcal S})\nonumber
\end{multline}
where the equality holds iff $\boldsymbol\Omega_2$ is a zero matrix,
i.e., iff ${\mathbf s}$ and ${\mathbf s}'$ are neighbouring points.
\hfill$\Box$

We can establish another useful property on the metric between
${\mathcal X}_{{\mathbf I}_M}({\mathbf s})$ and ${\mathcal
X}_{{\mathbf I}_M}({\mathbf s}')$ by first recalling an important
property in matrix algebra\cite{Lancaster}:

\noindent The characteristic polynomial of an $M\times M$ matrix
$\mathbf A$ is the polynomial whose roots are the eigenvalues of
$\mathbf A$. Mathematically, it can be re-written as
\begin{eqnarray}
    h(\nu) &\triangleq& \det~(\mathbf I + \nu\mathbf A)\nonumber\\
    &=& \nu^M + c_1\nu ^{M-1} + \cdots +\nonumber\\
    && c_{M-1}\nu + c_M \label{eq:charpolynomial}\\
    {\rm such~that}\qquad c_m &=& \sum_\vartheta \det (\mathbf
    A)_{i_1,\cdots,i_m} \label{eq:coefficients}
\end{eqnarray}
where $\mathbf A_{i_1,\cdots,i_m}$ denotes the principal submatrix
obtained by deleting the rows and columns of $\mathbf A$ except the
$i_1$th, the $i_2$th, $\cdots$, and the $i_m$th ones, and
$\vartheta$ denotes the combination set of $i_1,\cdots,i_m$. We
note, in particular, $c_1={\rm tr}~({\mathbf A})$ and
$c_M=\det(\mathbf A)$. Now, the following property provides us with
another lower bound on the metric between ${\mathcal X}_{{\mathbf
I}_M}({\mathbf s})$ and ${\mathcal X}_{{\mathbf I}_M}({\mathbf s}')$
in relation to $d_{\rm min}({\mathcal S})$.

\begin{property}\label{pro:lb}
Let ${\mathbf\Delta}={\rm diag}(\delta_1, \delta_2, \cdots,
\delta_M)$ with $\delta_m>0$ for $m=1, 2, \cdots, M$. 
Then, for any nonzero vector ${\mathbf e}$, the following inequality
holds
\begin{eqnarray}\label{eq:gfunction}
    \det\left({\mathbf\Delta}+{\mathcal X}_{{\mathbf I}_M}^H({\mathbf
    e}){\mathcal X}_{{\mathbf I}_M}({\mathbf
    e})\right)\ge\prod_{m=1}^M \left(\delta_m+d^2_{\rm min}({\mathcal
    S})\right)
\end{eqnarray}
with equality holding if and only if ${\mathbf s}$ and ${\mathbf
s}'$ are neighbours.~\hfill\QED
\end{property}
\noindent\textit{Proof
}: Let us first rewrite the left side of Eq.~(\ref{eq:gfunction}) as
\begin{multline}\label{eq:express}
    \det\left({\mathbf \Delta}+{\mathcal X}_{{\mathbf I}_M}^H({\mathbf
    e}){\mathcal X}_{{\mathbf I}_M}({\mathbf e})\right)=\\\det({\mathbf
    \Delta})\det\left({\mathbf I}+\left({\mathcal X}_{{\mathbf
    I}_M}({\mathbf e}){\mathbf\Delta}^{-1/2}\right)^H{\mathcal
    X}_{{\mathbf I}_M}({\mathbf e}){\mathbf\Delta}^{-1/2}\right)
\end{multline}
Now, let ${\mathbf A}=\left({\mathcal X}_{{\mathbf I}_M}({\mathbf
e}){\mathbf\Delta}^{-1/2}\right)^H{\mathcal X}_{{\mathbf
I}_M}({\mathbf e}){\mathbf\Delta}^{-1/2}$, then
Eqs.~(\ref{eq:charpolynomial}) and (\ref{eq:coefficients}) becomes
\begin{eqnarray}\label{eq:exp}
    \det\left[{\mathbf I}+\left({\mathcal X}_{{\mathbf I}_M}({\mathbf
    e}){\mathbf\Delta}^{-1/2}\right)^H{\mathcal X}_{{\mathbf
    I}_M}({\mathbf e}){\mathbf\Delta}^{-1/2}\right]=1+\sum_{m=1}^M
    c_m,
\end{eqnarray}
where
\begin{multline}\label{eq:coeffc_m}
\hspace{-.4cm} c_m=\sum_{\vartheta} \det\big[{\mathcal
X}^H_{{\mathbf
    I}_M}({\mathbf e},\{i_1,i_2,\cdots,i_m\})\\{\mathcal
    X}_{{\mathbf I}_M}({\mathbf e},\{i_1,i_2,\cdots,
    i_m\})\big]\prod_{\ell=1}^m \delta_{i_\ell}^{-1}
\end{multline}
Using Eq. (\ref{eq:detsubmatrix}) in Property~\ref{pro:submatrix} on
the right side of Eq.~(\ref{eq:coeffc_m}), we have
\begin{eqnarray}\label{eq:coefficient lb}
    c_m\ge d^{2m}_{\rm min}({\mathcal S})\sum_{\vartheta}
    \prod_{\ell=1}^m \delta_{i_\ell}^{-1}
\end{eqnarray}
where $\sum_{\vartheta}\prod_{\ell=1}^m \delta_{i_\ell}^{-1}$
denotes the sum of the combination of the product of
$\delta_{i_\ell} $taken $m$ at a time. Equality in
Eq.~(\ref{eq:coefficient lb}) holds if and only if ${\mathbf s}$ and
${\mathbf s}'$ are neighbours. Combining Eqs.~(\ref{eq:exp}) and
(\ref{eq:coefficient lb}) results in\vspace{-.3cm}
\begin{subequations}
\begin{eqnarray}
   && \det\left[{\mathbf I}+\left({\mathcal X}_{{\mathbf I}_M}({\mathbf
    e}){\mathbf\Delta}^{-1/2}\right)^H{\mathcal X}_{{\mathbf
    I}_M}({\mathbf e}){\mathbf\Delta}^{-1/2}\right]\nonumber\\&\ge&
    1+\sum_{m=1}^M d^{2m}_{\rm min}({\mathcal
    S})\sum_{\vartheta} \prod_{\ell=1}^m
    \delta_{i_\ell}^{-1}\label{eq:lower1}\\
    &=&\prod_{m=1}^M\big(1+\delta_m^{-1}d^{2}_{\rm min}({\mathcal
    S})\big)\label{eq:lower2}
\end{eqnarray}
\end{subequations}
where in the second step, we recognize that the right side of
Eq.~(\ref{eq:lower1}) is the eigen polynomial of the matrix
$d^{2}_{\rm min}\boldsymbol \Delta$ which, in turn, equals $\det
(\mathbf I + d^{2}_{\rm min}\boldsymbol \Delta)$ and hence
Eq.~(\ref{eq:lower2}). Combining Eqs.~(\ref{eq:lower2}) and
(\ref{eq:express}), we complete the proof of
Property~\ref{pro:lb}.~\hfill$\Box$

\section{Toeplitz STBC Applied to a MISO System with a Linear Receiver}\label{sec:ZFreceiver}
We now apply the Toeplitz STBC to the MISO communication system
using the properties presented in Section~\ref{sec3:properties}.
From Property~\ref{pro:full-rate} and \ref{pro:toeplitz lower bound}
in Section~\ref{sec3:properties} and
Theorem~\ref{theo:full-diversity} in
Section~\ref{sec:full-diversity}, we can see that the Toeplitz STBC
can approach \emph{unit-rate} as well as \emph{full diversity} even
if only a linear ZF or linear MMSE receiver is used in a MISO
system. In the following,
we examine the optimal tradeoff between diversity gain and
multiplexing gain~\cite{zheng03} when the Toeplitz STBC is employed
in a MISO system equipped with a linear receiver. We first make the
assumption that the channel coefficients are independent, i.e.,
${\mathbf\Sigma}= {\mathbf I}$.

Now, our MISO system has $M$ transmitter antennas transmitting a
signal vector $\mathbf s$ of length $L$ in $N=L+M-1$ time slots.
Also, all the above three signalling schemes have constellation
cardinality $\mu$. Thus, employing any of the three schemes
described in Section~\ref{sec:DesignCriterion} in our MISO system
will result in a transmission data rate $r$ given by\vspace{-.3cm}
\begin{eqnarray}\label{eq:rate_constellation1}
    r=\frac{L}{N}\log_2\mu
\end{eqnarray}
Note that $r$ is the \emph{bit} rate of transmission. The
multiplexing gain $g$, on the other hand, is dependent on the scheme
and in general, is defined as~\cite{zheng03}\vspace{-.3cm}
\begin{eqnarray}\label{eq:multiplexing}
    g_i=\frac {r}{\log_2{\rm SNR}_i}
\end{eqnarray}
where ``SNR$_i$" refers to the general SNR in the received data.
Here in our analysis, we use the SNR of the received \emph{data
block} for the SNR$_i$ and denote this by $\rho_{{\rm
bl}i},~i=1,~2,~3$ when the $i$th signalling scheme is employed.
Notice that in the MISO system, we always have $0\le g_i\le
1~\forall~i$, since the system has only one receiver antenna. Hence,
we can write $\mu=\rho_{{\rm bl}i}^{Ng_i/L}$. Thus, if we want to
maintain nonzero multiplexing gain at high SNR, then $\mu$, the
cardinality of the constellation must be large. It has been
shown~\cite{zheng03} that at high SNR, we can trade-off the
multiplexing gain for diversity gain and \emph{vice versa}, and the
optimum trade-off for our MISO system with $M$ transmitter antennas
is given by
\begin{equation}\label{eq:optradeoff}
    D_{\rm op} = M(1-g)
    \end{equation}
where $D_{\rm op}$ is the optimal diversity gain.
Let us examine trade-off of the multiplexing gain for diversity gain
in the three signalling schemes:
\begin{enumerate}[1.]
\item \emph{QAM signals}: The averaged symbol energy $E_s$ for
square QAM signal is~\cite{proakis00}
\begin{eqnarray}\label{eq:signal_power}
    E_{s1}=\frac{2}{3}(\mu-1)
\end{eqnarray}
We note that $E_{s1}$ increases with the constellation cardinality
$\mu$. From Eq.~(\ref{eq:signal_power}), the averaged transmission
energy \emph{per block} can be calculated as $E_{{\rm bl}1}=
\frac{2}{3}(\mu-1)ML$. Given $\sigma^2$ being the noise variance at
the receiver antenna, the averaged noise power per block is
$\sigma^2_{{\rm bl}1}=\sigma^2 N$. Therefore, the block SNR is
\begin{eqnarray}
    \rho_{{\rm bl}1}=\frac{2(\mu-1)ML}{3N\sigma^2}\nonumber
\end{eqnarray}
leading to
\begin{eqnarray}\label{eq:QAMnoise}
     \sigma^2 \approx\frac{2{\mu}ML}{3N\rho_{{\rm bl}1}}
        =\frac{2ML}{3N}{\rho_{{\rm bl}1}}^{\frac{Ng_1}{L}-1}
\end{eqnarray}
where the approximation is under the assumption of large $\mu$, and
Eq.~(\ref{eq:multiplexing}) has been used. Therefore, from
Eqs.~\eqref{eq:symbolSNR}, \eqref{eq:a_i}, \eqref{eq:signal_power}
and \eqref{eq:QAMnoise}, we obtain
\begin{eqnarray}\label{eq:multiplexity1}
    a_1\rho_1=\frac{3E_s}{4(\mu-1)\sigma^2}=\frac{1}{2\sigma^2}=
    \frac{3N}{4ML}~\rho_{{\rm bl}1}^{1-\frac{Ng}{L}}
\end{eqnarray}
Now, consider Eq.~\eqref{eq:average unification sep} on the upper
bound of the SEP for a ZF receiver, i.e.,
\begin{eqnarray}\label{eq:SEP_1}
    {\rm E}\left[{P}_1({\mathbf h})\right]\le C_{T\min}^{-M}(a_1\rho_1)^{-M}
\end{eqnarray}
Substituting Eq.~\eqref{eq:multiplexity1} into Eq.~\eqref{eq:SEP_1},
we obtain
\begin{eqnarray}\label{eq:QAMSEPbound}
    {\rm E}\left[{P}_1({\mathbf h})\right]\le
    C_{T\min}^{-M}\left(\frac{3N}{4ML}\right)^{-M}
    {\rho_{1\rm bl}}^{\frac{MN}{L}g-M}
\end{eqnarray}
Hence, the diversity gain for this scheme, $D_1$, is given by
\begin{multline}\label{eq:diversity_1}
    D_1(g)=M\left(1-\frac{N}{L}g\right)
    =M(1-g)-\varepsilon Mg\\=D_{\rm op}(g)-\varepsilon Mg
\end{multline}
where $\varepsilon=\frac{M-1}{L}\ge 0$. From
Eq.~\eqref{eq:diversity_1}, we can see that $D_1(g)\le D_{\rm
op}(g)$. However, we can make $\varepsilon$ small by choosing $L$
sufficiently large so that $D_1(g)\approx D_{\rm op}(g)$. Hence,
$D_1(g)$ is the $\varepsilon$-approximation of $D_{\rm op}(g)$. We
can always choose
    $L=\lceil\frac{M-1}{\varepsilon}\rceil+1$,
where $\lceil\cdot\rceil$ denotes the integer part of a quantity,
and therefore, we can say that \emph{the ZF receiver is able to
approach the optimal diversity-multiplexing tradeoff if the proposed
Toeplitz code is used with a square QAM signalling scheme of large
cardinality}.

If a MMSE receiver is employed, utilizing an expression parallel to
Eq.~(\ref{eq:QAMSEPbound}) for the MMSE receiver, we can also show
that the optimal diversity-multiplexing tradeoff can be
asymptotically achieved if the Toeplitz STBC is applied to the MISO
system in which a square QAM signalling scheme of large cardinality
is used for transmission.

\item \emph{PAM signals}: We note that the averaged transmission
energy $E_s$ for $\mu$-ary PAM signal is given by\cite{proakis00}
$E_s=\frac{1}{6}(\mu^2-1)$. Hence the averaged transmission energy
per block is $E_{s\rm bl}= \frac{1}{6}(\mu^2-1)ML$. Following
similar arguments resulting in Eq.~\eqref{eq:QAMnoise}, for PAM
signals we have,
\begin{eqnarray}
     \sigma^2
        \approx\frac{{\mu}^2ML}{6N\rho_{{\rm bl}2}}
        =\frac{ML}{6N}{\rho_{{\rm bl}2}}^{\frac{2Ng}{L}-1}\nonumber
\end{eqnarray}
Also, similarly to Eq.~\eqref{eq:multiplexity1}, we have
\begin{eqnarray}\nonumber
    a_2\rho_2=\frac{3E_s}{2(\mu^2-1)\sigma^2}=\frac{1}{4\sigma^{2}}=
    \frac{3N}{2ML}~{\rho_{{\rm bl}2}}^{1-\frac{2Ng}{L}}
\end{eqnarray}
Therefore, from Eq.~\eqref{eq:average unification sep}, the upper
bound on SEP for PAM signal is
\begin{eqnarray}\nonumber
    {\rm E}\left[{P}_2({\mathbf h})\right]&\le&
    C_{T\min}^{-M}(a_2\rho_2)^{-M}\nonumber\\
    &=&\left(\frac{3NC_{T\min}}{2ML}\right)^{-M}
    {\rho_{{\rm bl}2}}^{\frac{2MN}{L}g-M}\nonumber
\end{eqnarray}
Hence, the diversity order is
\begin{multline}\label{diversity_2}
    D_2(g)=M\left(1-\frac{2N}{L}g\right)=M(1-2g)-2\varepsilon Mg\\\le M(1-2g)
    \le M(1-g)= D_{\rm op}(g)
\end{multline}
Equality in \eqref{diversity_2} holds iff $g=0$. Therefore,
\emph{for finite multiplexing gain, $D_2(g)$ cannot approach the
optimal tradeoff $D_{\rm op}(g)$}.

\item \emph{PSK signals}: The averaged transmission energy $E_s$
for $\mu$-ary PSK signal is~\cite{proakis00} $E_s=1$. Hence the
averaged transmission energy per block is $E_{s\rm bl}=ML$.
Therefore, we have
\begin{eqnarray}
     \sigma^2
        =\frac{ML}{N\rho_{{\rm bl}3}}\nonumber
\end{eqnarray}
We also have
\begin{eqnarray}\label{multiplexity3}
    a_3\rho_3=\frac{E_s\sin^2(\pi/\mu)}{2\sigma^2}\approx\frac{\pi^2}{2\sigma^2\mu^2}=
    \frac{N\pi^2}{2ML}~{\rho_{{\rm bl}3}}^{1-\frac{2Ng}{L}}
\end{eqnarray}
where, the second step comes from the assumption of large $\mu$.
Following similar arguments as PAM scheme, it can be shown that
\emph{PSK signalling cannot achieve the optimal tradeoff of
diversity-multiplexing gains.}
\end{enumerate}

\section{Optimal Toeplitz STBC Design for MISO System with ML Detector\label{sec:ML}}
The previous section shows what could be achieved when the Toeplitz
STBC is applied to a MISO system equipped with a \emph{linear}
receiver. In this section, we will examine the application of the
Toeplitz STBC to a MISO system equipped with a ML detector. In
particular, we seek for the optimal design of the matrix ${\mathbf
B}$ inherent in a Toeplitz space-time block code
Eq.~(\ref{eq:toeplitz coding matrix}) such that the worst case
pair-wise error probability is minimized when a maximum likelihood
detector is employed.

Given a channel realization ${\mathbf h}$ and a transmission matrix
$\mathbf B$, the probability $P\left({\mathbf s}\rightarrow{\mathbf
s}'|{\mathbf h},{\mathbf B}\right)$ of transmitting ${\mathbf s}$
and deciding in favor of ${\mathbf s}'\ne {\mathbf s}$ with the ML
detector is given by~\cite{forney98}
\begin{eqnarray}\label{eq:pair}
    P\left({\mathbf s}\rightarrow{\mathbf s}'|{\mathbf
    h},{\mathbf B}\right)=Q\left(\frac{d({\mathbf s},{\mathbf
    s}')}{2\sigma}\right)
\end{eqnarray}
where $d({\mathbf s},{\mathbf s}')$ is the Euclidean distance
between the Toeplitz coded signals $\mathbf s$ and $\mathbf s'$
after being transmitted through the channel, i.e., it is the
Euclidean distance between ${\mathcal X}_{\mathbf B}({\mathbf
s}){\mathbf h}$ and ${\mathcal X}_{\mathbf B}({\mathbf s'}){\mathbf
h}$. Because of the relation of Eq.~(\ref{eq:toeplitz coding
matrix}), we can write:
\begin{eqnarray}
    d^2({\mathbf s},{\mathbf s}')&=&({\mathbf s}-{\mathbf s}')^H{\mathcal
    T}^H(\widetilde{\mathbf h},M,N){\mathcal T}(\widetilde{\mathbf
    h},M,N)\,({\mathbf s}-{\mathbf s}')\nonumber\\
    &=&{\mathbf h}^H{\mathcal X}^H_{\mathbf B}({\mathbf
    e}){\mathcal X}_{\mathbf B}({\mathbf e})\,{\mathbf h}
\end{eqnarray}
where ${\mathbf e}={\mathbf s}-{\mathbf s}'$. By employing the
alternative expression of the $Q$-function in Eq.~(\ref{eq:q}) and
taking the average of Eq.~(\ref{eq:pair}) over the Gaussian random
vector ${\mathbf h}$, the average pair-wise error probability can be
written as
\begin{multline}\label{eq:pairwise-error}
    P\left({\mathbf s}\rightarrow{\mathbf
    s}'|{\mathbf B}\right)=\frac{1}{\pi}\int_{0}^{\pi/2}\\\frac{d\theta}{{\rm
    det}\left({\mathbf
    I}+(8\sigma^2\sin^2\theta)^{-1}{\mathbf\Sigma}{\mathcal
    X}^H_{\mathbf B}({\mathbf e}){\mathcal X}_{\mathbf B}({\mathbf
    e})\right)}
\end{multline}
with $\boldsymbol\Sigma$ being the covariance matrix of $\mathbf h$.
Our design problem can now be stated as:

\noindent\emph{Design Problem}: For a fixed number $M$ of
transmitter antennas, find a $K\times M$, ($K\le M$), matrix
${\mathbf B}$ such that the worst-case average pair-wise error
probability $P({\mathbf s}\rightarrow{\mathbf s}'|{\mathbf B})$ is
minimized, subject to the transmission power constraint, ${\rm
tr}\left({\mathbf B}^H{\mathbf B}\right)\le 1$, i.e.,
\begin{eqnarray}\label{eq:designB}
    {\mathbf B}_{\rm op}={\rm arg}\;\min\limits_{{\rm tr}({\mathbf
    B}^H{\mathbf B})\le 1}~~\max_{\begin{subarray}{c}
    \mathbf{s},~ {\mathbf s}'\in {\mathcal S}^L\\
    {\mathbf s}'\ne {\mathbf s}'
    \end {subarray}}
    P({\mathbf s}\rightarrow{\mathbf s}'|{\mathbf B})
\end{eqnarray}
where ${\mathcal S}^L = {\mathcal S}\times {\mathcal
S}\cdots\times{\mathcal S}$ with ${\mathcal S}$ denoting the
signal constellation of each element of ${\mathbf s}$.

To solve the above design problem, we not only have to find the
optimum $\mathbf B$, but also have to determine its dimension $K$.
Let us first examine the $M\times M$ covariance matrix, $\boldsymbol
\Sigma = {\rm E}[\mathbf{hh}^H]$, of the transmission channels
$\mathbf h$. Suppose we perform an eigen decomposition such that
$\boldsymbol \Sigma = {\mathbf V}{\mathbf \Lambda}{\mathbf V}^H$
where ${\mathbf V}$ is an $M\times M$ unitary matrix and ${\mathbf
\Lambda}={\rm diag}(\lambda_1, \lambda_2, \cdots, \lambda_M)$ with
$\lambda_1\ge\lambda_2\ge\cdots\ge\lambda_M>0$. The following
theorem provides us with an optimum design of $\mathbf
B$:\vspace{-.2cm}
\begin{theorem}\label{th:Bop}
Let $\mathbf\Gamma={\rm diag}(\gamma_1, \gamma_2, \cdots,
\gamma_K),~K\le M,$ be the singular values of $\mathbf B$ and let
$G({\mathbf\Lambda_K}{\mathbf\Gamma}, \varepsilon)$ denote the
integral\vspace{-.2cm}
\begin{eqnarray}\label{eq:G}
    G({\mathbf\Lambda}_K{\mathbf\Gamma},
    \varepsilon)=\frac{1}{\pi}\int_{0}^{\pi/2}\prod_{k=1}^K
    \left(1+\frac{\varepsilon\lambda_k\gamma_k^2}{\sin^2\theta}\right)^{-1}d\theta
    \quad\textrm{for}~\varepsilon>0
\end{eqnarray}
We obtain an optimal $\mathbf\Gamma$ by solving the following convex
optimization problem:\vspace{-.1cm}\footnote{ Note that the work
presented here is different from that in~\cite{Hehn} in which a
precoder matrix is designed for a frequency-selective fading channel
even though both involve Toeplitz structured matrices. Here, the
Toeplitz matrix containing $\mathbf B$ and $\mathbf h$ is separated
from the signal vector $\mathbf s$. This, by the properties of the
Toeplitz STBC shown, transforms the design of $\mathbf B$ into a
convex optimization problem. In \cite{Hehn}, however, the design
parameter and the signal vector are all parts of the Toeplitz
structure resulting in a non-convex design problem that can only be
solved by numerical method with no guarantee for global optimality.}
\begin{eqnarray}\label{eq:OpGamma}
    {\boldsymbol\Gamma}_{\rm op}={\rm arg}\;\min\limits_{{\rm
    tr}({\boldsymbol\Gamma})\le
    1}G\left({\boldsymbol\Lambda}_K{\boldsymbol\Gamma}, \frac{d^2_{\rm
    min}({\mathcal S})}{8\sigma^2}\right)
\end{eqnarray}
where ${\boldsymbol\Gamma}_{\rm op}$ is a $K\times K$ diagonal
matrix given by ${\boldsymbol\Gamma}_{\rm op}={\rm
diag}(\gamma_{{\rm op}1},~\gamma_{{\rm op}2},~\cdots,~\gamma_{{\rm
op}K})$ with $K$ being the highest integer for which $[{\mathbf
\Gamma}_{\rm op}]_{kk}={\gamma}_{{\rm op}k}>0$, $k=1, 2, \cdots, K$.
Then the optimum transmission matrix is given by\vspace{-.1cm}
\begin{equation}\label{optB}
    {\mathbf B}_{\rm op}=\mathbf \Gamma_{\rm op}{\mathbf V}_K^H
\end{equation}
where $\mathbf V_K$ is the $M\times K$ matrix containing the $K$
eigenvectors corresponding to the $K$ largest eigenvalues in the
eigen decomposition of $\mathbf\Sigma$. Furthermore, the worst case
pair-wise error probability is lower bounded by
\begin{eqnarray}\label{eq:lowerbound}
        \max\limits_{{\mathbf s},{\mathbf s}'\in {\mathcal S}^L, {\mathbf s}
        \ne {\mathbf s}'} P({\mathbf s}\rightarrow
        {\mathbf s}'|\mathbf B)\ge G\left({\boldsymbol\Lambda}_K{\boldsymbol\Gamma}_{\rm
        opt},\,\frac{d^2_{\rm min}({\mathcal S})}{8\sigma^2}\right)
\end{eqnarray}
Equality in Eq.~(\ref{eq:lowerbound}) holds if and only if
\begin{enumerate}[i)]
\item $\|{\mathbf s}-{\mathbf s}'\|=d_{\rm min}({\mathcal S})$,
and \item $\mathbf B= \mathbf B_{\rm op}$.~\hfill\QED
\end{enumerate}
\end{theorem}
{\textit {Proof}}: We first establish a lower bound on the worst
case average pair-wise error probability. Let $\mathbf s_\ell$ and
$\mathbf s'_\ell$ be neighbour symbol vectors differing in only the
$\ell$th symbol, i.e., $\mathbf s_\ell - \mathbf s_\ell'=\mathbf
e_\ell = [0~\cdots~0~~e_\ell~~0~\cdots~0]^T$ where $|e_\ell|=d_{\rm
min}({\mathcal S})$. Then, we can write
\begin{eqnarray}\nonumber
    &&{\rm det}\big({\mathbf
    I}_M+\frac{1}{8\sigma^2\sin^2\theta}{\boldsymbol\Sigma}{\mathcal
    X}^H_{{\mathbf B}}({\mathbf e_\ell}){\mathcal X}_{{\mathbf B}}({\mathbf
    e_\ell})\big)\nonumber\\
    &&={\rm det}\big({\mathbf I}_M+\frac{d^2_{\rm min}({\mathcal
    S})}{8\sigma^2\sin^2\theta}{\boldsymbol\Sigma}{\mathbf B}^H{\mathbf B}\big)\nonumber\\
    &&\le\prod_{k=1}^K\big(1+\frac{d^2_{\rm min}({\mathcal
    S})}{8\sigma^2\sin^2\theta}\gamma_k^2\lambda_k\big)\label{eq:hadamard}
\end{eqnarray}
where the first step is a result of the structure of $\mathbf
e_\ell$ on the Toeplitz code, and the second step is the result of
an inequality for the determinant of a matrix~\cite{witsenhausen75,
cover91}. Equality in Eq.~(\ref{eq:hadamard}) holds if and only if
${\mathbf B}={\mathbf B}_o={\mathbf \Gamma}{\mathbf V_K}^H$, i.e.,
the singular vectors of $\mathbf B$ are the eigenvectors of
$\boldsymbol\Sigma$. Substituting the inequality of
(\ref{eq:hadamard}) in Eq.~(\ref{eq:pairwise-error}), we have
$P({\mathbf s_\ell}\rightarrow{\mathbf s_\ell}'|\mathbf B)\ge
G\left({\mathbf \Lambda}_K{\mathbf \Gamma},\,\frac{d^2_{\rm
min}({\mathcal S})}{8\sigma^2}\right)$. Since
$\left(\max\limits_{{\mathbf s},{\mathbf s}'\in {\mathcal S}^L,
{\mathbf s}\ne {\mathbf s}'} P({\mathbf s}\rightarrow{\mathbf
s}'|\mathbf B)\right)\ge P({\mathbf s_\ell}\rightarrow{\mathbf
s_\ell}'|\mathbf B)$, the worst case average pair-wise error
probability is lower bounded by
\begin{eqnarray}\label{eq:low-bound}
    \max\limits_{{\mathbf s}, {\mathbf s}'\in {\mathcal S}^L, {\mathbf
    s}\ne {\mathbf s}'}P({\mathbf s}\rightarrow{\mathbf
    s}'|{\mathbf B})
    \ge G\left({\mathbf \Lambda}_K{\mathbf \Gamma},\,\frac{d^2_{\rm min}({\mathcal
    S})}{8\sigma^2}\right)
\end{eqnarray}
If we minimize both sides of Eq.~(\ref{eq:low-bound}), we can write
\begin{equation}\label{eq:lower-bound}
    \min_{\mathbf B}\left(\max\limits_{{\mathbf s}, {\mathbf s}'
    \in {\mathcal S}^L, {\mathbf
    s}\ne {\mathbf s}'}P({\mathbf s}\rightarrow{\mathbf
    s}'|{\mathbf B})\right)\ge G\left({\mathbf \Lambda}_K{\mathbf \Gamma}_{\rm op},
    \,\frac{d^2_{\rm min}({\mathcal S})}{8\sigma^2}\right)
\end{equation}
where $\mathbf \Gamma_{\rm op}$ is obtained according to
Eq.~(\ref{eq:OpGamma}).

Let us now establish an upper bound for the worst case average
pair-wise error probability for the specially structured
transmission matrix $\mathbf B_o$ above. For any error vector
$\mathbf e$, we have
\begin{multline}\label{eq:Gequality}
    {\rm det}\!\!\left(\!{\mathbf
    I}_M \!\! +\!\! \frac{1}{8\sigma^2\sin^2\theta}{\boldsymbol\Sigma}{\mathcal
    X}^H_{{\mathbf B_o}}({\mathbf e}){\mathcal X}_{{\mathbf B_o}}({\mathbf
    e})\!\right)
    \!\\=\!\left(\frac{1}{8\sigma^2\sin^2\theta}\right)^{M}\!\!{\rm
    det}\!({\boldsymbol\Lambda}_K{\mathbf \Gamma}^2) {\rm
    det}\!\left({\mathbf\Delta}\!+\!{\mathcal X}^H_{{\mathbf I}_M}({\mathbf
    e}){\mathcal X}_{{\mathbf I}_M}({\mathbf e})\right)
\end{multline}
where the special structure of $\mathbf B_o$ has been utilized,
$\boldsymbol\Lambda_K$ denotes the diagonal matrix containing the
largest $K$ positive eigenvalues of $\boldsymbol\Sigma$ and
${\mathbf\Delta}=(8\sigma^2\sin^2\theta){\boldsymbol\Lambda}_K^{-1}{\mathbf
\Gamma}^{-2}$. Using Eq.~(\ref{eq:Gequality}) in
Property~\ref{pro:lb}, for any nonzero vector ${\mathbf e}$ and
nonzero $\theta$ in the interval $[0, \pi/2]$, we have
\begin{eqnarray}\label{eq:Ginequality}
    {\rm det}\big({\mathbf\Delta}+{\mathcal X}^H_{{\mathbf
    I}_M}({\mathbf e}){\mathcal X}_{{\mathbf I}_M}({\mathbf e})\big) \ge
    \prod_{k=1}^M\big(\frac{8\sigma^2\sin^2\theta}{\gamma_k^2\lambda_k}+d^2_{\rm
    min}({\mathcal S})\big)
\end{eqnarray}
where, according to Property~\ref{pro:lb}, equality holds if and
only if ${\mathbf s}$ and ${\mathbf s}'$ are neighbour vectors.
Eq.~(\ref{eq:Gequality}) and Eq.~(\ref{eq:Ginequality}) together
yield
\begin{multline}\label{eq:GinequalityA}
     {\rm det}\big({\mathbf
    I}_M+\frac{1}{8\sigma^2\sin^2\theta}{\boldsymbol\Sigma}{\mathcal X}^H_{{\mathbf
    B_o}}({\mathbf e}){\mathcal X}_{{\mathbf B_o}}({\mathbf
    e})\big)
    \ge\\\prod_{k=1}^M\big(1+\frac{d^2_{\rm min}({\mathcal
    S})\gamma_k^2\lambda_k}{8\sigma^2\sin^2\theta}\big)
\end{multline}
Again, substituting Eq.~(\ref{eq:GinequalityA}) in
Eq.~(\ref{eq:pairwise-error}) and using the optimum
$\mathbf\Gamma_{\rm op}$ yields
\begin{equation}
    \max\limits_{{\mathbf s}, {\mathbf s}'\in {\mathcal S}^L,\,{\mathbf
    s}\ne {\mathbf s}'}P({\mathbf s}\rightarrow{\mathbf s}'|{{\mathbf B_{\rm op}}})\le
    G\left({\boldsymbol\Lambda}_K{\boldsymbol\Gamma}_{\rm op},\,\frac{d^2_{\rm
    min}({\mathcal S})}{8\sigma^2}\right)\nonumber
\end{equation}
where equality holds if and only if $\|{\mathbf s}-{\mathbf
s}'\|=d_{\rm min}({\mathcal S})$. This results in
\begin{eqnarray}
    &&\min_{\mathbf B}\left(\max\limits_{{\mathbf s}, {\mathbf s}'\in {\mathcal S}^L,\,{\mathbf
    s}\ne {\mathbf s}'}P({\mathbf s}\rightarrow{\mathbf s}'|{{\mathbf
    B}})\right)\nonumber\\
    &&\le\max\limits_{{\mathbf s}, {\mathbf s}'\in {\mathcal S}^L,\,{\mathbf
    s}\ne {\mathbf s}'}P({\mathbf s}\rightarrow{\mathbf s}'|{{\mathbf B_{\rm op}}})\nonumber\\
    &&\le
    G\left({\boldsymbol\Lambda}_K{\boldsymbol\Gamma}_{\rm op},\,\frac{d^2_{\rm
    min}({\mathcal S})}{8\sigma^2}\right)\label{eq:upper}
\end{eqnarray}
Combining Eq.~(\ref{eq:upper}) with Eq.~(\ref{eq:lower-bound})
yields
\begin{eqnarray}\label{eq:min-bound}
    \min_{\mathbf B}\left(\max_{\begin{subarray}{c} {\mathbf s}, {\mathbf s}'\in {\mathcal S}^L\\
    {\mathbf s}\ne {\mathbf s}'
    \end {subarray}}P({\mathbf s}\rightarrow{\mathbf
    s}'|{{\mathbf B}})\right)=
    G\left({\boldsymbol\Lambda}_K{\boldsymbol\Gamma}_{\rm op},\,\frac{d^2_{\rm
    min}({\mathcal S})}{8\sigma^2}\right)
\end{eqnarray}
Eq.~(\ref{eq:min-bound}) holds iff ${\mathbf B}={\mathbf
\Gamma}_{\rm op}{\mathbf V}_K^H$ and $\|{\mathbf s}-{\mathbf
s}'\|=d_{\rm min}({\mathcal S})$. Thus, the proof of
Theorem~\ref{th:Bop} is complete.\hfill$\Box$

\noindent\underline{Remarks on Theorem~\ref{th:Bop}}:
\begin{enumerate}[a)]
\item Theorem~\ref{th:Bop} shows us that the lower bound of the
worst case pair-wise error probability can be reached by having
$\mathbf B= {\mathbf\Gamma}_{\rm op}{\mathbf V}_K^H$. Thus, the
design problem in Eq.~(\ref{eq:designB}) becomes finding the optimum
$\mathbf\Gamma_{{\rm op}}$ of Eq.~(\ref{eq:OpGamma}).
\item The original non-convex optimization problem has been
transformed into the convex problem in Eq.~(\ref{eq:OpGamma}) and
can be solved efficiently by interior point methods. The convexity
of the objective function can be verified by re-writing
$G({\mathbf\Lambda}_K{\mathbf\Gamma},\varepsilon)$ in
Eq.~\eqref{eq:G} as
\begin{multline}\label{eq:G_cov}
    G({\mathbf\Lambda}_K{\mathbf\Gamma},
    \varepsilon)=\frac{1}{\pi}\int_{0}^{\pi/2}\hspace{-.4cm}\exp\left(-\sum_{k=1}^{K}\ln
    \left(1+\frac{\varepsilon\lambda_k\gamma_k^2}{\sin^2\theta}\right)
    \right)d\theta,\\\varepsilon>0
\end{multline}
We notice that $-\ln(\cdot)$ is a convex function over $\gamma_k^2$,
and hence their sum is also convex over
$[\gamma_1^2,\cdots,\gamma_K^2]$. Now, $\exp(x)$ is monotonically
increasing with $x$. By \emph{composition rule}~\cite{convex}~(Page
84), the integrand in Eq.~\eqref{eq:G_cov} is a convex function
implying that $G({\mathbf\Lambda}_K{\mathbf\Gamma},\varepsilon)$ is
convex.

\item The solution of Eq.~(\ref{eq:OpGamma}) yields the values of
the diagonal elements $\{\gamma_{{\rm op}1},~\gamma_{{\rm
op}2},~\cdots,~\gamma_{{\rm op}M}\}$. Some of these values may not
be positive. We choose all the $K$ positive ones to form the
singular values of $\mathbf B_{\rm op}$.
\end{enumerate}

\noindent Theorem~\ref{th:Bop} provides us with an efficient scheme
to obtain the optimal matrix $\mathbf B_{\rm op}$ by numerically
minimizing $G({\mathbf\Lambda}_K{\mathbf\Gamma},\varepsilon)$.
However, if the Chernoff bound \cite{simon98} of the pairwise error
probability is employed as the objective function for minimization
instead, a closed-form optimal $\mathbf B$ can be obtained. This can
be shown by setting $\sin^2\theta=1$ in the pairwise error
probability of Eq.~\eqref{eq:pairwise-error}, so that we obtain the
Chernoff bound as \vspace{-.2cm}
\begin{eqnarray}\label{eq:Chernoff-bound}
    P\left({\mathbf s}\rightarrow{\mathbf
    s}'|{\mathbf B}\right)\le\frac{1}{2{\rm
    det}\left({\mathbf
    I}+(8\sigma^2)^{-1}{\mathbf\Sigma}{\mathcal
    X}^H_{\mathbf B}({\mathbf e}){\mathcal X}_{\mathbf B}({\mathbf
    e})\right)}
\end{eqnarray}
Seeking to minimize the worst case Chernoff bound, and following
similar arguments which establish the the optimization problem in
Eq. \eqref{eq:OpGamma}, we arrive at the following problem,
\begin{eqnarray}\label{relax}
    \widetilde{\boldsymbol \Gamma}_{\rm op}={\rm arg}\;\min\limits_{{\rm
    tr}(\widetilde{\boldsymbol \Gamma})\le
    1}\frac{1}{2}\prod_{k=1}^K
    \left(1+\frac{d^2_{\rm min}({\mathcal
    S})}{8\sigma^2}\lambda_k\widetilde\gamma_k^2\right)^{-1}
\end{eqnarray}
where $\widetilde{\boldsymbol \Gamma}_{\rm op}$ is a diagonal matrix
with diagonal elements $\widetilde\gamma_{{\rm op}k}$. This problem
is a relaxed form of that in Eq.~(\ref{eq:OpGamma}) and its solution
is provided by the following corollary:

\begin{corollary}\label{corollary:water-filling}
The solution, $\widetilde{\mathbf \Gamma}_{\rm op}$, for the
optimization problem of Eq.~\eqref{relax} can be obtained by
employing the water-filling strategy~\cite{cover91}. The diagonal
elements of $\widetilde{\mathbf \Gamma}_{\rm op}$ are given by
\begin{multline}\label{water}
    \widetilde{\gamma}_{{\rm op}k }=\sqrt{\left[\frac{1}{M_0}
    \left(1+\frac{8\sigma^2}{d^2_{\rm min}({\mathcal S})}
    \sum_{\ell=1}^{M_0}\frac{1}{\lambda}_{\ell}\right)-
    \frac{1}{\lambda_n}\right]_{+}},\\ k = 1,\cdots, K
 \end{multline}
where notation $[x]_{+}$ denotes ${\rm max}(x,0)$. The optimal
choice of $K$ is $K=M_0$, where $M_0$ is the maximum positive
integer satisfying
\begin{eqnarray}\nonumber
    \frac{1}{M_0}\left(1+\frac{8\sigma^2}{d^2_{\rm min}({\mathcal S})}
    \sum_{\ell=1}^{M_0}\frac{1}{\lambda}_{\ell}\right)-\frac{1}{\lambda_m}>0,
    \quad m=1, 2, \cdots, M_0
\end{eqnarray}
The optimum transmission matrix is thus $\widetilde{\mathbf B}_{\rm
op} =\widetilde{\mathbf \Gamma}_{\rm op}{\mathbf V}_K^H$.
\end{corollary}
\emph{Proof}: The minimization of the type of problems of
Eq.~\eqref{relax} has been studied by several researchers and the
water-filling solution is shown in~\cite{cover91}.~\hfill$\Box$

\noindent\underline{Remarks on
Corollary~\ref{corollary:water-filling}}:
\begin{enumerate}[a)]
    \item For the particular case in which the channel coefficients
    are mutually independent, i.e., ${\boldsymbol\Sigma}={\mathbf
    I}_M$, then any $M\times M$ unitary matrix scaled by a factor
    $1/\sqrt{M}$ is a suitable choice for $\widetilde{\mathbf B}_{\rm
    op}$.
    \item The optimal design $\widetilde{\mathbf B}_{\rm op}$
    maximizes the \emph{coding gain} \cite{tarokh98} which is defined
    to be the normalized minimum determinant of $\mathbf{\mathcal
    X}(\mathbf e)^H\mathbf{\mathcal X}(\mathbf e)$ for all nonzero
    $\mathbf e$. The criterion of coding gain maximization is derived
    from minimizing the Chernoff bound on pairwise error probability
    for a system equipped with a ML detector. Since
    $\widetilde{\mathbf B}_{\rm op}$ minimizes the worst case
    Chernoff bound on PEP, we conclude that it achieves optimal coding
    gain.
\end{enumerate}
\noindent\underline{Remarks on the Optimum Transmission Matrix
Design}:
\begin{itemize}
    \item The derivations of Theorem~\ref{th:Bop} and
    Corollary~\ref{corollary:water-filling}
    are based on the consideration of using a ML receiver in the MISO system.
    For the case when the system is equipped with a linear ZF (or MMSE) receiver under the
    environment of correlated channels, the problem of obtaining an optimum $\mathbf B$
    becomes very complicated. This is because we seek for a matrix $\mathbf B$ to minimize
    the respective average error probability obtained by averaging the expressions of error
    probability in Eqs. \eqref{eq:qamsep}, \eqref{eq:pamsep} and
    \eqref{eq:psksep} for the respective signalling schemes over the random channel matrix.
    This requires the knowledge of the PDF of the equivalent channel matrix. However, the
    equivalent channel matrix in these cases is of a Toeplitz structure for which the PDF
    is unavailable, and therefore, the expressions for the average error probability cannot
    be obtained. (For the case of linear MMSE receivers, a similar problem
    exists).

    \item An alternative way to attack the problem in the case when a linear receiver
    is employed is to consider the upper bound on the averaged error probability
    given in Eq.~\eqref{eq:average unification sep}. We can minimize this upper bound with
    respect to $\mathbf B$. However, this necessitates the knowledge of the value of $C_0$.
    From Property~\ref{pro:toeplitz lower bound}, $C_0$ is the minimum value of the
    determinant of the Toeplitz matrix having its column vector belonging to unit ball
    and this renders the solving of $C_0$ difficult. Thus, in this paper, we
    are unable to come up with any true optimal $\mathbf B$ for MISO systems equipped with linear receivers.
\end{itemize}


\section{Numerical Experiments\label{sec7}}
In this section, we examine the performance of the Toeplitz code in
a MISO system. We first evaluate the performance of the system
equipped with a linear receiver under the condition that different
parameters are varied. We then evaluate the performance of the
system employing different beamformers as well as a linear or a ML
receiver in an environment in which the channels are correlated. We
note that for a linear receiver, the major computation occurs in the
inversion of the Toeplitz matrix for which the complexity is of
order $\mathcal O(LM)$~\cite{Golub83}, where $L$ is the length of
signal vector and $M$ is the number of transmitter antennas for the
MISO system. On the other hand, the complexity using a ML detector
for this MISO system transmitting the Toeplitz code is of order
$\mu^M$ where $\mu$ is the constellation cardinality. Thus, for a
reasonably large constellation and/or a comparatively large number
of transmitter antennas, the ML receiver is substantially more
complex than a linear receiver. Finally, we compare the performance
of the Toeplitz code with some other known efficient codes.
\begin{figure}
    \centering \resizebox{7cm}{!}
    {\includegraphics{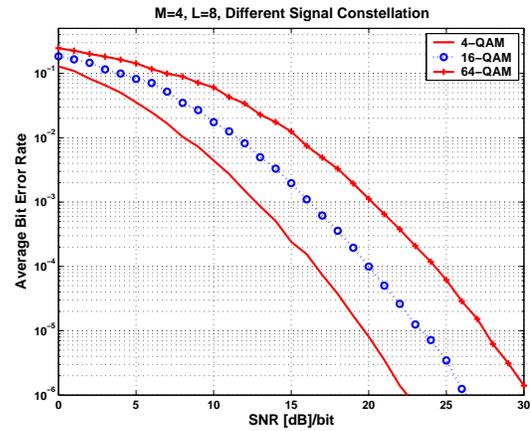}}
    \begin{center}
        \caption{The average BER performance of the proposed Toeplitz STBC
        when signals are selected from different constellations.}\vspace{-.3cm}
        \label{fig:constellation}
    \end{center}
\end{figure}
\begin{figure}
    \centering \resizebox{7cm}{!}
    {\includegraphics{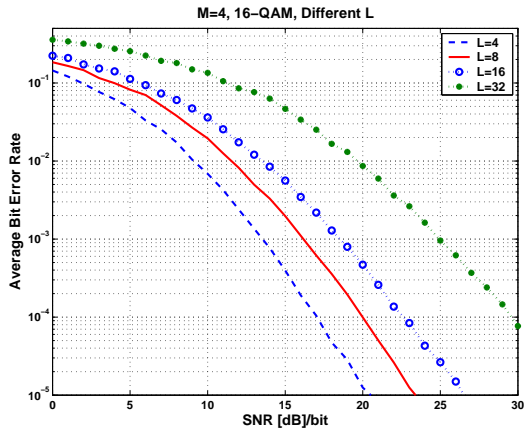}}
    \begin{center}
        \caption{The average BER performance of the proposed Toeplitz STBC
        for different $L$.}\vspace{-.3cm}\label{fig:L}
    \end{center}
\end{figure}
\begin{figure}
    \centering \resizebox{7cm}{!}
    {\includegraphics{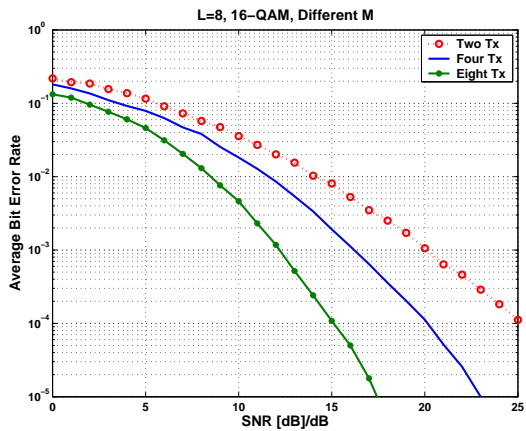}}
    \begin{center}
        \caption{The average BER performance of the proposed Toeplitz STBC
        for the MISO system with different number of transmitter antennas.}\vspace{-.3cm}
        \label{fig:M}
    \end{center}
\end{figure}

\noindent{\bf Example 1}: In this example, we examine the
performance of Toeplitz STBC for a MISO communication system with
independent channel fading, i.e., $\boldsymbol\Sigma=\mathbf I$. The
system 
is equipped with a linear ZF detector at the receiver end. For the
Toeplitz STBC, we choose $\mathbf B=\mathbf I$ in
Eq.~\eqref{eq:toeplitz coding matrix}. The following three
experiments are performed:

\begin{enumerate}
    \item We fix the number of transmitter antennas to be $M=4$
     and the length of signal vector $\boldsymbol s$ to be $L=8$,
    and the symbol transmission data rate is therefore
    $R_s=L/N=0.7273$ symbols pcu. The signals are randomly selected
    from the different constellations of 4-QAM, 16-QAM and 64-QAM. The
    signals are transmitted through the MISO system having zero-mean
    unit-variance i.i.d. Gaussian channels and additive white Gaussian noise as
    described in Section~\ref{sec:DesignCriterion} and the BER curves are plotted
    in Fig.~\ref{fig:constellation} in which the three different curves correspond
    to the performance of the system using the three signal constellations respectively.
    It should be noted that different constellation size results in different transmission
    bit rates. For the system we examine, the bit rates are $R_b=1.4545$, $2.9091$, and
    $5.8182$ bits pcu corresponding to 4-QAM, 16-QAM and 64-QAM respectively. Therefore,
    for larger constellation size, worse BER performance is expected. This is indeed
    the case as shown by the three BER curves plotted in
    Fig.~\ref{fig:constellation} from which it is observed that, for a BER of $10^{-5}$,
    the difference in SNR between 4-QAM and 16-QAM is approximately 3dB and that between
    16-QAM and 64-QAM is approximately 5 dB.

    \item In this experiment, we fix the signal constellation to be 16-QAM
    for a four transmitter antenna MISO system and
    perform simulations for different signal lengths, $L=4,~8,~16,~32$.
    For the different choices of $L$, the system has different
    transmission symbol rates, which are $R_s=0.5714,~0.7273,~0.8421,
    ~0.9143$ symbols pcu, respectively. The channel and noise assumptions are the same
    as those in the previous experiment. The BER curves at different SNR are plotted in
    Fig.~\ref{fig:L}. It can be observed that the longer is the transmitted signal,
    the worse is the system BER performance. Again, this is due to the fact
    that larger $L$ corresponds to higher transmission date rate resulting in
    worse performance.

    \item In this experiment, we vary the number of the transmitter
    antennas $M$. An increase in $M$ increases the diversity and
    decreases the transmission symbol rate.
    Therefore, it is expected that the performance of the system will be
    enhanced with the increase of the number of transmitter antennas.
    This is indeed the case as illustrated in Fig.~\ref{fig:M} where we
    compare the performance of the MISO systems having $M= 2,4$, and 8 antennas
    with $L=8$ in the Toeplitz codes and the signals selected from a 16-QAM
    constellation.
\end{enumerate}

\noindent{\bf Example 2}: In this example, we test the performance
of Toeplitz STBC for correlated channels in a MISO system equipped
with four transmitter antennas in a linear array and the one
receiver antenna on the normal to the axis of the transmitter
antenna array (``broadside"). For small angle spread, the
correlation coefficient between the $m_1$th and $m_2$th transmitter
antennas is\cite{shiu, zhou-it03}
\begin{eqnarray}\label{correlation}
    [\boldsymbol\Sigma]_{m_1m_2}\approx\frac{1}{2\pi}
    \int_{0}^{2\pi}\exp\left(
    -j2\pi
    (m_1-m_2)\Delta\frac{d_t}{\varsigma}\sin\theta\right)d\theta
\end{eqnarray}
where $d_t$ is the antenna spacing, $\varsigma$ is the wavelength of
the (narrowband) signal, and $\Delta$ is the angle spread. Here in
our simulations, we choose $d_t=0.5\varsigma$ and $\Delta=5^\circ$.
We examine the performance of the MISO system transmitting 4-QAM
signals in the following three cases using Toeplitz STBC having the
structures:
\begin{enumerate}[i)]
\item $\mathbf B=\frac{1}{\sqrt{M}}\mathbf I$. This is an
approximately optimal transmission matrix at sufficiently high SNR
in the minimization of the Chernoff bound under the assumption that
$\mathbf B$ is a square matrix. The approximate optimality can shown
as follows: For Eq.~\eqref{eq:Chernoff-bound}, under high SNR, we
ignore the identity matrix $\mathbf I$ in the denominator and obtain
\begin{eqnarray}\label{eq:Chernoff-bound_relax}
   \hspace{-.4cm} P\left({\mathbf s}\rightarrow{\mathbf
    s}'|{\mathbf B}\right)\hspace{-.2cm}&<&\hspace{-.2cm}\frac{(8\sigma^2)^M}{2~{\rm
    det}\left({\mathbf\Sigma}{\mathcal
    X}^H_{\mathbf B}({\mathbf e}){\mathcal X}_{\mathbf B}({\mathbf
    e})\right)}\nonumber\\
    \hspace{-.2cm}&=&\hspace{-.2cm}\frac{(8\sigma^2)^M}{2~
    \det({\mathbf\Sigma})\det\left(\mathbf B^H{\mathcal
    T}^H({\mathbf e}){\mathcal T}({\mathbf e})\mathbf B\right)}
\end{eqnarray}
To minimize the right side of Eq.~\eqref{eq:Chernoff-bound_relax},
we maximize the second determinant in the denominator. We note that

\begin{eqnarray}
    &&\nonumber\det\left(\mathbf B^H{\mathcal
    T}^H({\mathbf e}){\mathcal T}({\mathbf e})\mathbf B\right)\\
    &&\le\det\left({\mathcal
    T}^H({\mathbf e}){\mathcal T}({\mathbf e})\right)\prod_{m=1}^{M}\left[\mathbf B\mathbf
    B^H\right]_{mm}\nonumber\\
    &&\le \det\left({\mathcal
    T}^H({\mathbf e}){\mathcal T}({\mathbf e})\right)
    \left(\frac{1}{M}{\rm tr}\left(\mathbf B\mathbf
    B^H\right)\right)^M\nonumber\\
    &&=\frac{1}{M^M}\det\left({\mathcal
    T}^H({\mathbf e}){\mathcal T}({\mathbf e})\right)\label{eq:relax}
\end{eqnarray}
The first inequality is due to Hadamard Inequality \cite{horn1985}
and equality holds iff $\mathbf B\mathbf B^H$ is diagonal. The
second inequality is due to geometric mean being no larger than
arithmetic mean, with equality holding iff $\mathbf B\mathbf B^H$
has equal diagonal elements. Finally, the trace of $\mathbf B\mathbf
B^H$ is equal to unity due to the power constraint. Hence, the
condition for maximum in Eq.~\eqref{eq:relax} is that $\mathbf B$ is
a scaled unitary matrix of which $\mathbf B= \mathbf I$ is one
choice.

\item $\mathbf B=\mathbf B_{\rm op}$. This is the optimal solution
to minimize the worst case PEP derived in Theorem~\ref{th:Bop} and
can be obtained numerically by solving the convex optimization
problem of Eq.~(\ref{eq:OpGamma}) and then using the result in
Eq.~(\ref{optB}).

\item $\mathbf B=\widetilde{\mathbf B}_{\rm op}$. This minimizes
Chernoff bound on PEP as described in
Corollary~\ref{corollary:water-filling}.
\end{enumerate}
In our simulations here, the transmitted signal vector is of length
$L=10$ and each of the symbols is randomly selected from the 4-QAM
constellation. At the receiver, the signals in each of the three
cases are detected separately by a linear ZF detector and a ML
detector and the respective performances of the two detectors are
examined. 
(As mentioned in the end of the last section, we cannot obtain an
exact optimum $\mathbf B$ for the ZF receiver. Nonetheless, we will
employ the two optimum transmission matrices derived for the ML
receiver $\mathbf B_{\rm op}$ and $\widetilde{\mathbf B}_{\rm op}$
to the case of ZF receiver to see how the performance is improved).
We note that due to the variation of the channel fading, the
dimension $K$ of the optimum transmission matrices in Cases ii) and
iii) change with SNR. For the specific correlated channel described
in Eq.~\eqref{correlation}, for Case ii), we found that $K=1$ when
SNR $\le$ 8dB and $K=2$ at higher SNR, whereas for Case iii), we
found that $K=1$ when SNR $\le$ 10dB and $K=2$ at higher SNR.
Therefore, the transmission data rate for these two cases is
$R_1=\frac{K}{L+K-1}$ symbols pcu, which is lower than
$R_2=\frac{M}{L+M-1}$ for the case of $\mathbf B=\mathbf I_M$. For a
fair comparison, we choose two different structures for $\mathbf B$
in Case i) in the following experiments:
\begin{enumerate}
    \item We maintain the transmission data rate in Case i) the same as
    that in case iii). This is realized by setting $\mathbf B =
    \big[~\frac{1}{\sqrt{K}}\mathbf I_K, \mathbf 0_{(K,M-K)}\big]$ in Case~i) with $K$
    being the dimension of $\widetilde{\mathbf B}_{\rm
    op}^H\widetilde{\mathbf B}_{\rm op}$. We evaluated the error
    performances of the systems equipped with different $\mathbf B$ in all
    three cases and the results are shown in Fig.~\ref{example2} from which the
    following observations can be made:

\begin{figure}
    \centering \resizebox{7cm}{!}
    {\includegraphics{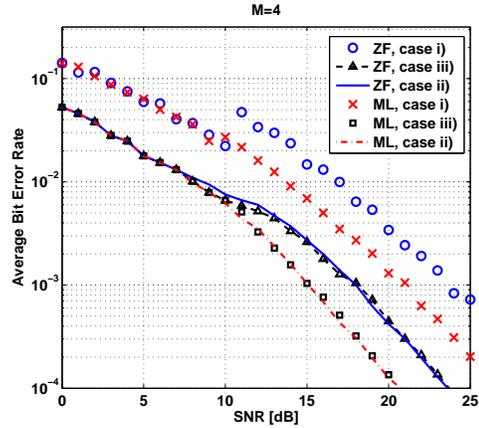}}
    \begin{center}
        \caption{The average bit error rate comparison of the proposed Toeplitz STBC
        with i) $\mathbf B=[\mathbf I_K,\mathbf 0_{M-K}]$, ii) $\mathbf B_{\rm op}$ and iii)
        $\widetilde{\mathbf B}_{\rm op}$.
        The performances are shown for both linear ZF detectors and ML detectors.}\vspace{-.3cm}
        \label{example2}
    \end{center}
\end{figure}
\begin{figure}
    \centering \resizebox{7cm}{!}
    {\includegraphics{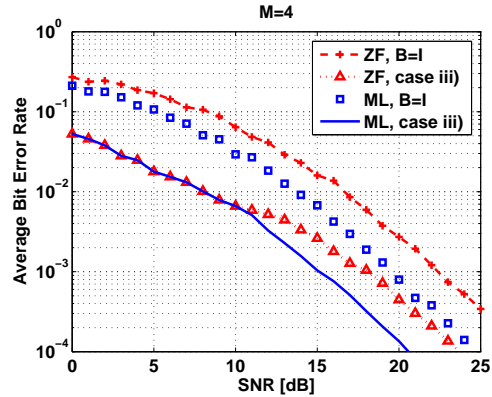}}
    \begin{center}
        \caption{The average bit error rate comparison of the proposed Toeplitz STBC
        with $\mathbf B=\mathbf I_M$ and $\widetilde{\mathbf B}_{\rm op}$.
        The performances are shown for both linear ZF detectors and ML detectors.}\vspace{-.3cm}
        \label{example3}
    \end{center}
\end{figure}
    \begin{itemize}
        \item For the system employing a ML detector, performance of Case ii) and iii)
        are superior to that of Case i), confirming the theoretical analyses in
        Theorem~\ref{th:Bop} and
        Corollary~\ref{corollary:water-filling}.

        \item For the system employing a ML detector, the BER performance
        for Cases ii) and iii) employing $\mathbf B_{\rm op}$ and
        $\widetilde{\mathbf B}_{\rm op}$ respectively are very close.
        This shows that Chernoff bound is tight for this system.
        Close performance in the two cases is also true for the
        system using a ZF detector.

        \item Although $\mathbf B_{\rm op}$ and $\widetilde{\mathbf B}_{\rm op}$
        are optimal transmission matrices developed for the ML detector,
        they are equally effective in providing substantial performance improvement
        for the same system employing a linear ZF detector.

        \item At lower SNR, we have $K=1$, i.e., only one transmitter
        antenna is effective. Therefore, given a coded system, linear ZF
        and ML detectors provide the same performance.
    \end{itemize}
    \item In the second part of the experiment, we put $\mathbf B=\mathbf I_M$
    in Case i) and examine its performance. The error performance for such a
    choice is shown in Fig.~\ref{example3}.
    Here, the system using $\mathbf B=\mathbf I_M$ has higher
    transmission data rate than those in Cases ii) and iii). For the sake of
    comparison, we have re-plotted in Fig.~\ref{example3} the performance
    curves from Fig.~\ref{example2} of Case iii) corresponding to the uses of
    $\widetilde{\mathbf B}_{\rm op}$ as a transmission matrix. (Since the
    performance of Cases ii) is almost the same as that of Case iii), we have
    omitted here the performance curves corresponding to the use of
    $\mathbf B_{\rm op}$).
    It should be noted that when the signals are detected by a ML
    detector, the system coded with $\mathbf B =\mathbf I_M$ has higher
    diversity gain over the system with $\widetilde{\mathbf B}_{\rm
    op}$. This is due to the fact that water-filling strategy may not
    employ all the available transmitter antennas for correlated
    channels. Specifically in this example, the effective number of
    antennas for $\widetilde{\mathbf B}_{\rm op}$ is $K \le 2<4$.
    However, the optimal coding gain achieved by $\widetilde{\mathbf
    B}_{\rm op}$ with ML detectors ensures a better performance. It is
    also important to note that the employments of $\mathbf B=\mathbf
    I_M$ and $\widetilde{\mathbf B}_{\rm op}$ result in a relatively
    large difference in performances, revealing that the upper bound on
    PEP given in Eq.~\eqref{eq:Chernoff-bound_relax} is not tight.
    Thus, even though this bound is quite commonly employed in STBC
    designs for independent channels, the results here show that this
    relaxed bound is a poor design criterion for an environment of
    highly correlated channel coefficients.
\end{enumerate}

\begin{figure}
    \centering \resizebox{7cm}{!}
    {\includegraphics{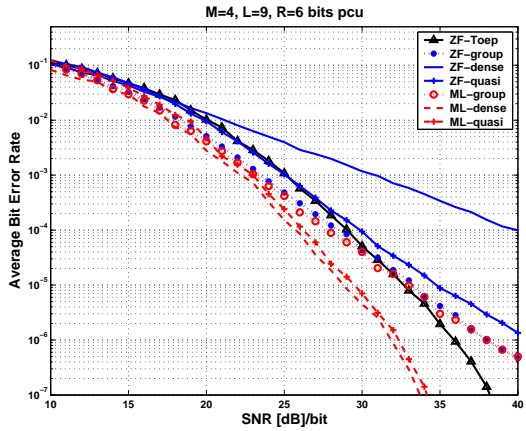}}
    \begin{center}
        \caption{The average bit error rate comparison of the proposed Toeplitz STBC
        with other STBC of unit rate.}\vspace{-.3cm}\label{fig:STBC_4t1r}
    \end{center}
\end{figure}

\noindent{\bf Example 3}: In this example, we compare the BER
 performance of Toeplitz STBC with other STBC for independent MISO
 channels. Here again, we choose $\mathbf B=\mathbf I$ for Toeplitz STBC. The
 experiments are performed for the two cases in which the number of
 transmitter antennas in the communication system are $M=4$ and $M=8$
 respectively:
\begin{enumerate}
\item $M=4$ transmitter antennas and a single receiver antenna: We
compare BER performance of Toeplitz STBC with other rate one STBC
\cite{quasi-xia, dense, quasi-dao, group}:
\begin{itemize}
    \item Quasi-orthogonal STBC. The code for four transmitter antennas
    was presented in~\cite{quasi-xia}, and the maximization of its coding gain
    was subsequently shown in~\cite{quasi-dao}.
    \item Dense full-diversity STBC \cite{dense}
    \item Multi-group decodable STBC \cite{group}
\end{itemize}
For the Toeplitz STBC, we choose $L=9$ for which the symbol
transmission data rate is $R_s=L/N=3/4$ symbols pcu. To achieve a
fair comparison, the same transmission \emph{bit} rate is imposed on
all the codes such that signals are selected from 256-QAM
constellation for Toeplitz STBC and from 64-QAM for the other
full-rate STBC. Therefore, the same transmission bit rate, $R_b=6$
bits pcu, is employed for all the systems. At the receiver, the
Toeplitz STBC is processed by a linear ZF equalizer followed by a
symbol-by-symbol detector. For the other full-rate STBC, we examine
the two cases in which the signals are processed by a) a ML detector
and b) a linear ZF receiver. The BER curves are plotted in
Fig.~\ref{fig:STBC_4t1r}. When a linear ZF equalizer and a
symbol-by-symbol detector is applied at the receiver, it can be
observed that Toeplitz STBC outperforms ``quasi-orthogonal" STBC and
``dense" STBC, and at higher SNR, its performance is superior to
multi-group code. It is also interesting to observe that  at higher
SNR, for Toeplitz STBC with linear ZF receivers, the performance is
also superior to that of the Multi-group STBC using a ML receiver.
In fact, for the range of SNR tested, the slope of its BER curve is
the same as those of the ``dense" STBC and the ``quasi-orthogonal"
STBC processed by ML detectors, indicating they have the same
diversity gain.
\begin{figure}
    \centering \resizebox{7cm}{!}
    {\includegraphics{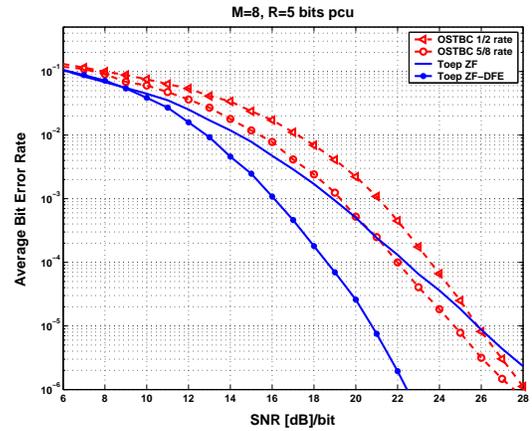}}
    \begin{center}
        \caption{The average bit error rate comparison of the proposed Toeplitz STBC
        with the orthogonal STBC.}\vspace{-.3cm}\label{fig:orthogonal}
    \end{center}
\end{figure}
\item We now consider the system having $M=8$ transmitter
antennas. For the Toeplitz code, we choose $L=35$ and therefore, the
symbol transmission data rate is $L/N=5/6$ symbols pcu. We compare
the bit error rate performance of our Toeplitz code with that of the
orthogonal STBC having symbol
transmission rate of:\\
i) $1/2$ symbols pcu~\cite{tarokh99,alamouti98,tirkkonen02} and\\
ii) $5/8$ symbols pcu \cite{ostbc-xia} (this the highest symbol rate
achievable by the orthogonal STBC applied to an eight transmitter
antenna system).\\
To achieve a fair comparison, the transmitted signals are selected
from a $64$-QAM constellation for our Toeplitz code, a $256$-QAM
constellation for the $5/8$ rate orthogonal code and a $1024$-QAM
constellation for the $1/2$ rate orthogonal code. Hence, all of the
codes have the same transmission data rate in bits, i.e., $R_b=5$
bits pcu. At the receiver end, the orthogonal STBC is decoded by a
linear ZF detector for which, because of the orthogonality, the
performance is the same as that of a ML detector. For Toeplitz STBC,
the signals are decoded separately by a linear ZF receiver and a
ZF-DFE receiver. 
The average bit error rate for these codes are plotted
Fig.~\ref{fig:orthogonal}. It can be observed that the performance
of the Toeplitz code detected with a linear ZF receiver is superior
to that of the $\frac{1}{2}$-rate orthogonal STBC when the SNR is
less than or equal to 25 dB. 
When the Toeplitz STBC is received by a ZF-DFE receiver, due to the
higher coding gain, its performance is significantly better than
that of the orthogonal STBC. In Fig.~\ref{fig:orthogonal} at
$10^{-5}$, the Toeplitz code with a ZF-DFE receiver outperforms the
orthogonal code by about 4 dB.

It should be noted that for the Toeplitz code, both linear ZF and
ZF-DFE receivers can achieve full-diversity.
However, from Fig.~\ref{fig:orthogonal}, while the slope of BER
curve for Toeplitz code with ZF-DFE receiver is similar to those of
the orthogonal codes, the slope of the curve for the Toeplitz code
with linear ZF receiver is not as steep. Recall that the diversity
gain of a communication system is defined at \emph{high} SNR and
here, the upper end of the SNR range is not sufficiently high. To
show full diversity for both systems, we need BER at higher SNR, the
evaluation of which demands exorbitant computation for the
parameters in this example. To circumvent this difficulty, we choose
to compare the \emph{symbol error rate} (SER) obtained by the use of
the Toeplitz code with linear ZF receiver to that
obtained by the use of the $5/8$ orthogonal code. 
The results are shown in Fig.~\ref{fig:ser} from which it can be
observed that the two SER curves have the same slope for SNR above
$30$dB, indicating the same diversity gain for both codes. Thus, we
can see that the Toeplitz code with a linear ZF (or more
sophisticated) receiver indeed achieves full diversity.
\begin{figure}
    \centering \resizebox{7cm}{!}
    {\includegraphics{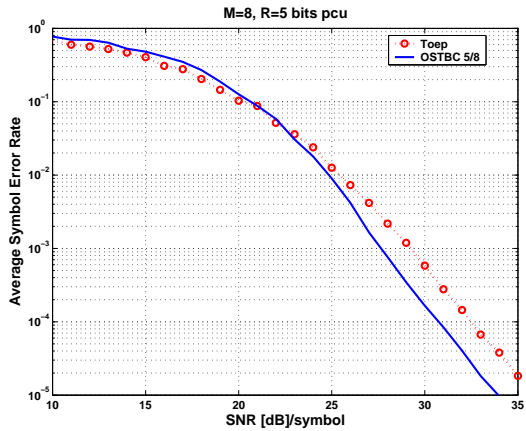}}
    \begin{center}
        \caption{The average symbol error rate comparison of the proposed Toeplitz STBC
        with the orthogonal STBC.}\vspace{-.3cm}\label{fig:ser}
    \end{center}
\end{figure}
\end{enumerate}


\section{Conclusion \label{sec5} }
In this paper, we have presented a general design criterion for
full-diversity linear STBC when the signals are transmitted through
a MISO communication system and processed by a linear receiver. This
is, to our knowledge, the first design criterion for linear
receivers to achieve full diversity. Specifically, we proposed a
linear Toeplitz STBC for a MISO channel which satisfies the
criterion and achieves full-diversity. We have shown that such a
code possesses many interesting properties, two of which
recapitulated here are of practical importance:
\begin{enumerate}
    \item The symbol transmission rate for the code approaches one when the number of
    channel uses ($N>M$) is large.

    \item If the signalling scheme has a constellation for which
    the distance between the nearest neighbours is nonzero
    (such as QAM),
    then employing the Toeplitz code results in a non-vanishing determinant.
\end{enumerate}

When employed in a MISO system equipped with a linear receiver (ZF
or MMSE), the Toeplitz code can provide full diversity. Furthermore,
when the number of channel uses is large, in an independent MISO
flat fading environment, the Toeplitz code can approach the
Zheng-Tse optimal diversity-multiplexing tradeoff.

When employed in a MISO system equipped with a ML detector, for both
independent and correlated channel coefficients, we can design the
transmission matrix inherent in the proposed Toeplitz STBC to
minimize the exact worst case average pair-wise error probability
resulting in full diversity and optimal coding gain being achieved.
In particular, when the design criterion of the worst case average
pair-wise error probability is approximated by the Chernoff bound,
we obtain a closed-form optimal solution.

The use of the Toeplitz STBC (having an identity transmission
matrix) in a MISO system fitted with a ZF receiver has been shown by
simulations to have the same slope of the BER curves to other full
rate STBC employing a ML detector, whereas even better performance
can be achieved by using receivers (such as ZF-DFE) more
sophisticated than the linear ones to detect the Toeplitz code. For
correlated channels, employing the optimum transmission matrices in
the Toeplitz code results in substantial additional improvements in
performance to using the identity transmission matrix. This
substantial improvement of performance is observed in either case
for which a ML or a ZF receiver is used.

\end{document}